\documentclass[conference]{IEEEtran}
\IEEEoverridecommandlockouts
\usepackage{verbatim}
\usepackage{setspace}
\usepackage{paralist}
\usepackage{mdwmath}
\usepackage{amssymb}
\usepackage{amsmath}
\usepackage{amsthm}
\usepackage{xfrac}
\usepackage{mathtools}
\usepackage{chngcntr}
\usepackage{tabulary}
\usepackage{booktabs}
\usepackage{epsfig}
\usepackage{wrapfig}
\usepackage{lipsum}
\usepackage{caption}
\usepackage{subcaption}
\usepackage{multirow}
\usepackage{algcompatible}
\usepackage{footnote}
\usepackage{paralist}
\usepackage[draft]{hyperref}
\usepackage{gensymb}

\usepackage{cite}
\usepackage{graphicx}
\usepackage{textcomp}
\usepackage[dvipsnames,table,xcdraw]{xcolor}
\def\BibTeX{{\rm B\kern-.05em{\sc i\kern-.025em b}\kern-.08em
    T\kern-.1667em\lower.7ex\hbox{E}\kern-.125emX}}
\usepackage{fancyhdr}

\newenvironment{myitemize}
{
   \vspace{0mm}
    \begin{list}{$\bullet$ }{}
        \setlength{\topsep}{0em}
        \setlength{\parskip}{0pt}
        \setlength{\partopsep}{0pt}
        \setlength{\parsep}{0pt}
        \setlength{\itemsep}{1mm}
}
{
    \end{list}
}

\title{Not All GPUs Are Created Equal: Characterizing Variability in Large-Scale, Accelerator-Rich Systems\vspace{-1ex}}

\author{\IEEEauthorblockN{Prasoon Sinha,
Akhil Guliani, Rutwik Jain, Brandon Tran, Matthew D. Sinclair and
Shivaram Venkataraman}
\IEEEauthorblockA{Computer Sciences Department, University of Wisconsin-Madison\\
Madison, United States of America\\
Email: \{psinha9, guliani, rnjain, bqtran2\}@wisc.edu,
\{sinclair, shivaram\}@cs.wisc.edu}}

\begin{document}

\maketitle
\begin{abstract}
  
Scientists are increasingly exploring and utilizing the massive parallelism of general-purpose
  accelerators such as GPUs for scientific breakthroughs.  As a result, datacenters, hyperscalers,
  national computing centers, and supercomputers have procured hardware to support this evolving
  application paradigm.  These systems contain hundreds to tens of thousands of accelerators, enabling peta- and exa-scale levels of compute for scientific workloads.  Recent work demonstrated that power
  management (PM) can impact application performance in CPU-based HPC systems, even when machines
  have the same architecture and SKU (stock keeping unit).  This variation occurs due to manufacturing variability and
  the chip's PM.  However, while modern HPC systems widely employ accelerators such as GPUs, it is unclear how much this variability affects applications.  Accordingly, we seek to characterize
  the extent of variation due to GPU PM in modern HPC and supercomputing systems.  We study a
  variety of applications that stress different GPU components on five large-scale computing
  centers with modern GPUs: Oak Ridge's Summit, Sandia's Vortex, TACC's Frontera and Longhorn, and
  Livermore's Corona. These clusters use a variety of cooling methods and GPU vendors.  In total, we
  collect over 18,800 hours of data across more than 90\% of the GPUs in these clusters.
  Regardless of the application, cluster, GPU vendor, and cooling method, our results show
  significant variation: 8\% (max 22\%) average performance variation even though the GPU
  architecture and vendor SKU are identical within each cluster, with outliers up to \textbf{1.5$\times$} slower than the median GPU.  These results highlight the
  difficulty in efficiently using existing GPU clusters for modern HPC and scientific workloads, and
  the need to embrace variability in future accelerator-based systems.
\end{abstract}

\noindent

\begin{IEEEkeywords}
Accelerator Architectures; Dynamic Voltage Scaling; Power Measurement; Temperature Measurement; Time Measurement
\end{IEEEkeywords}

\vspace{-1ex}
\section{Introduction}
\label{sec:intro}

Recently, domain scientists have leveraged the massive parallelism of accelerators for scientific discovery.
Some of these discoveries use machine learning (ML) or deep learning (DL) for image recognition~\cite{Deng2009, krizhevsky2012imagenet}, speech recognition~\cite{AmodeiAnubhai2016-deepSpeech2, hannun2014deep, he2018streaming}, and machine translation~\cite{ott2018scaling, WuSchuster16-seq2seq}.
Scientists have also used accelerators in other areas, including molecular dynamics, material science, and quantum chemistry.
Since these applications often require peta- or exascale levels of compute, running them on massively parallel systems has yielded promising results in areas including protein folding~\cite{jumper2021highly}, plasma reactor status prediction~\cite{kates2019predicting}, material science~\cite{fan2021predicting}, and SARS-CoV-2~\cite{casalino2021ai}.

Accordingly, supercomputers, datacenters, and computing centers have procured hardware to accommodate this evolving application paradigm.
To reach exascale levels of compute, many of these systems use many accelerators, which offer greater power efficiency and thus support emerging AI and HPC workloads within a constrained power budget.
For example, nearly all of the top 10 supercomputers leverage GPUs~\cite{top500} and the second-ranked supercomputer, Oak Ridge National Lab's (ORNL's) Summit, has over 27000 GPUs.
Compute centers such as NCSA Delta (840 GPUs), SDSC Expanse (216 GPUs), and Texas Advanced Computing Center (TACC, 744 GPUs) also utilize many GPUs.
Similarly, Microsoft, Tesla, and others have deployed accelerator-based supercomputers~\cite{FowersOvtcharov2018-brainwave, tesla}.
The upcoming Aurora, El Capitan, and Frontier supercomputers~\cite{Frontier} are expected to have even more GPUs.
Thus, current exascale computing systems contain hundreds to tens of thousands of GPUs, and future systems will likely be comprised of a large variety of accelerators and customized chips~\cite{AbtsRoss2020-tsp, FowersOvtcharov2018-brainwave, TPU2, JouppiYoung2017-tpu, JouppiYoon2021-tpuv4, la2020cerebras, lie2021multi, mohan2020studying, tesla, ZheTillman2019-graphcore}.

Despite their power efficiency, exascale systems have an enormous power footprint,
making it important to consider the hardware's \emph{power management} (PM) algorithms.
PM can lead to power and frequency variations across nodes.
Such dynamic behavior makes it challenging for %
repeatable, high performance and can lead to resource underutilization.
For example, recent work studying CPU-based supercomputers showed that PM impacts application performance by up to 20\%, even for CPUs with the same architecture and vendor SKU (Stock-Keeping Unit)~\cite{AcunLanger2016-power,chasapis2016runtime, ChasapisMoreto2019-powerEfficJobSched, InadomiPatki2015-scVar, PatelWagenhauser2020-hpcPowerConsump, SkinnerKramer2005-perfVarCauses}. This variation occurs due to the manufacturing process and the chip's power constraints~\cite{ChasapisMoreto2019-powerEfficJobSched,Scogland2015-pwrPerspectives}.
However, despite their increasingly widespread use in modern HPC systems, there is limited work that
examines how accelerator PM and manufacturing variability affects application performance.
For example, while Scogland, et al.~\cite{Scogland2015-pwrPerspectives} conducted an AMD GPU variability study in 2015, they only used one benchmark and called for a more in-depth study.
We discuss related work further in Section~\ref{sec:related}.

\begin{figure}[tb!]
  \includegraphics[width=\columnwidth]{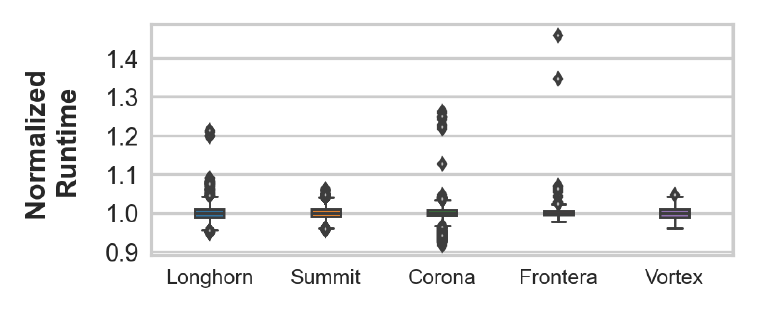}
  \vspace{-0.7cm}
  \caption{Normalized runtime across the five compute clusters for SGEMM. All clusters exhibit significant performance variability
  and contain several outliers.}
  \label{fig:sgemm-cluster-summary}
  \vspace{-4ex}
\end{figure}

In this paper we perform a rigorous study to understand GPU variability in large scale, accelerator-rich computing clusters.
While some systems~\cite{FowersOvtcharov2018-brainwave, tesla} utilize FPGAs or other accelerators, we focus on GPUs since the majority of the Top-500 systems utilize GPUs. %
To ensure coverage across GPU workloads that are run on modern systems and to stress different GPU components, we select five applications from ML, sparse graph analytics, and molecular dynamics.
Furthermore, we study multiple computing clusters to examine how different scales (hundreds to tens of thousands of GPUs), cooling (air, mineral oil, and water), and GPU vendors (AMD and NVIDIA) impact variability.
Specifically, we study five modern compute clusters: ORNL's Summit, Sandia National Lab's (SNL's) Vortex, TACC's Frontera \& Longhorn, and Lawrence Livermore National Lab's (LLNL's) Corona. We perform measurements on over 90\% of the GPUs on each cluster and verify that our methodology is statistically significant~\cite{Scogland2014-MeasurementLevels}.
In total, we collected and analyzed over 18,800 hours of data to study variability in performance, temperature, GPU frequency, and power consumption.
Our findings include:

\begin{myitemize}
  \vspace{-1ex}
  \item %
  All clusters exhibit significant performance variability when running SGEMM (Figure~\ref{fig:sgemm-cluster-summary}), which is normalized to a median runtime of 1. %
 On average, SGEMM performance varies between 7\%-9\% across GPUs in the same cluster, with outliers up to \textbf{1.5$\times$} slower than the median GPU.
  \item Performance variability is larger for compute intensive workloads such as matrix multiplication (SGEMM) or training ML models like ResNet-50 (22\%, max \textbf{3.5$\times$} slower than median GPU) than memory intensive workloads like PageRank (1\%). BERT pre-training, which has a mix of compute and memory-bound operations, has 8\% performance variation. This variation impacts resource utilization, increases runtime, and hurts responsiveness, especially for multi-GPU experiments.
  \item Air cooled clusters (Longhorn, Corona) have a large temperature range $\geq$ 30$\degree C$ and very high temperature nodes can have severe
      power throttling ($>$15$\%$ of TDP).  Water and mineral oil cooling reduce temperature variation, but do not reduce performance and power variation.
  \item On the largest GPU cluster we study (Summit), there are a number of power outliers that are not correlated with temperature or frequency variation.%
  \item Our results are consistent across different days of the week and times of day, indicating that the variability is not transient. Moreover, we also validated that similar or higher variability occurs when limiting the GPU power to be lower than the TDP (Section~\ref{subsec:res-dayOfWeek}).
  \vspace{-1ex}
\end{myitemize}

Overall our results indicate that significant power and performance variation exists in GPU-rich systems.
Given there are no standards to expose PM information to administrators, it is difficult today to identify aberrations within the cluster that might require maintenance, modifications, or even replacement of GPUs.
Further, unlike CPU-based systems, GPU nodes cannot be assigned as a desktop or a submit node to minimize impact. %
Thus, our results motivate systematic benchmarking across nodes to provide an early-warning for system administrators to perform maintenance or investigate bad GPUs, without hurting long-term cluster performance.
Finally, our work also highlights the need to improve visibility of accelerator variability and design mitigation techniques in runtime frameworks for both users and system administrators.

\section{Background and Related Work}
\label{sec:related}

\subsection{Hardware Performance Variability}
\label{subsec:related-hwPerfVar}

As large scale datacenters and supercomputers are typically power-constrained, manufacturing variability between two identical architectures translates to varying performance and causes load imbalance even for perfectly balanced workloads.
PM and temperature variations can also cause thermal throttling, possibly degrading performance further.
To mitigate variability, CPU-based HPC systems have adopted various techniques including: 
dynamic load balancing algorithms~\cite{AcunMiller2016-variationTurbo, AcunKale2016-varLoadBal},
cooling mechanisms and temperature-aware job placement~\cite{AcunLee2017-proactiveCooling, MenonAcun2013-thermAwareLoadBal,ZhangOgrenci2015-minThermalVar}, 
and intelligent adaptive runtimes~\cite{AcunLanger2016-power,chasapis2016runtime,ChasapisMoreto2019-powerEfficJobSched}.

However, less work has examined how GPU PM and manufacturing variability affect applications.
Coplin, et al.~\cite{CoplinBurtscher2016-gpgpuPower} and Jiao, et
al.~\cite{JiaoLin2010-gpuPowerPerf} demonstrated that memory- and compute-bound GPU
workloads exhibit significant differences in performance and energy for a variety of configurations.
Jiao, et al. also found that increasing frequency often increased both power
consumption and performance in Fermi-class GPUs.
Moreover, they demonstrated that GPUs utilize dynamic voltage and frequency scaling (DVFS) to help stay within their power limit.
However, both studies were conducted on GPUs several generations older than the ones in use today and focus on performance variability for a single-GPU workstation.
Similarly, Scogland, et al. examined how LINPACK's performance varied for CPU clusters and AMD GPUs~\cite{Scogland2015-pwrPerspectives}.
However, this study was done in 2015, only used a single benchmark, and motivates the need for a more in-depth study of GPU variability.
Thus, while these papers are a useful foundation, characterization is needed to determine the impact of PM on modern GPU-rich systems for representative workloads.

Other recent work has also examined how the Titan supercomputer's thousands of GPUs behaved over the machine's lifetime~\cite{ostrouchov2020gpulifetimes}.
However, this work focused on issues like reliability, and did not investigate the impact of PM.

\vspace{-1ex}
\subsection{GPU Power Management}
\label{subsec:related-gpuPM}

Current many-GPU systems use a local-only PM setup: each GPU has a given thermal limit it must stay within (its thermal design power, TDP), which is 300W for NVIDIA V100 GPUs and AMD MI60 GPUs~\cite{volta,mi60} and 230W for the NVIDIA Quadro RTX 5000 GPUs~\cite{rtx5000} we study. %
The GPU PM controller varies the GPU's voltage and frequency using DVFS to avoid exceeding its TDP~\cite{BharadwajDas2022-gpuDVFS, GeVogt2013-dvfsKepler}.
Although GPU vendors such as AMD and NVIDIA have not disclosed details about their DVFS schemes (or PM controllers), prior work has shown that, similar to multi-core CPUs, DVFS adjusts the GPU's streaming multiprocessor (SM) and memory frequencies and voltages to stay within the TDP and reduce energy consumption.
Sometimes DVFS inadvertently compromises performance~\cite{BharadwajDas2022-gpuDVFS, GeVogt2013-dvfsKepler, MeinerzhagenTokunaga2018-gpuDVFS}.
Thus, PM can affect the performance of GPU applications.
Moreover, there is no global PM strategy across GPUs in a cluster.
As a result, GPUs can be also affected by the behavior of other GPUs on the same node.

\begin{table}[bt!]
    {\footnotesize
  \centering
  \begin{tabular}{|l|c|c|c|c|}
    \hline
    \textbf{Cluster} & \textbf{GPU} & \textbf{\# GPUs} & \textbf{\# Nodes} & \textbf{Cooling} \\ \hline
    \textbf{CloudLab~\cite{DuplyakinRicci2019-cloudlab}} & NVIDIA V100 & 12 & 4 & air \\ \hline
    \textbf{Longhorn}~\cite{tacc} & V100 & 416 & 104 & air \\ \hline
    \textbf{Frontera}~\cite{tacc} & RTX 5000 & 360 & 90 & mineral oil \\ \hline
    \textbf{Vortex}~\cite{snl} & V100 & 216 & 54 & water \\ \hline
    \textbf{Summit}~\cite{ornl} & V100 & 27648 & 4608 & water \\ \hline
    \textbf{Corona}~\cite{llnl} & MI60 & 328 & 82 & air \\ \hline
  \end{tabular}
  \vspace{-1ex}
  \caption{Summary of clusters studied.}
  \label{tab:summary-hpc-clusters}
}

    \vspace{-4ex}
\end{table}

\vspace{-1ex}
\section{Methodology}
\label{sec:method}

To ensure that our results were representative across a variety of GPU clusters with different properties, we sought to run experiments across clusters of different sizes, cooling approaches, and GPU vendors.
Moreover, we also selected benchmarks that are representative of how modern systems are used and which stress a variety of GPU components.

\noindent
\textbf{Cluster Parameters}: Table~\ref{tab:summary-hpc-clusters} summarizes the unique HPC clusters we studied.
First, we used a small cluster in CloudLab~\cite{DuplyakinRicci2019-cloudlab}, which contains 12 air-cooled GPUs split across 3 nodes, each with 4 NVIDIA V100-SXM2 GPUs.
We also studied five larger HPC clusters.
TACC's air-cooled Longhorn cluster~\cite{tacc} has 416 GPUs split across 104 nodes, each with 4 NVIDIA V100 GPUs.
SNL's water-cooled Vortex cluster~\cite{snl} contains 216 NVIDIA V100 GPUs split across 54 nodes.
To examine variability in a larger cluster, we also studied ORNL's water-cooled Summit supercomputer~\cite{ornl} (27648 NVIDIA V100 GPUs).
Finally, to examine variability across GPU vendors, we studied AMD MI60 GPUs in Livermore's Corona cluster~\cite{llnl}.
Although we do not have access to Vortex's temperature setup, the Summit and TACC GPU's shutdown, slowdown, max operating, and max memory operating temperatures are: 90\degree C, 87\degree C, 83\degree C, and 85\degree C, respectively.  The Corona GPU's shutdown, slowdown, and max memory operating temperatures: are 105\degree C, 100\degree C, and 99\degree C, respectively~\cite{amd-power, vega11-maxTemp}.  Finally, the Frontera GPU's shutdown, slowdown, and max operating temperatures are: 96\degree C, 93\degree C, and 89\degree C, respectively.  Throughout our experiments, we did not observe GPUs exceed these thresholds.

\noindent
\textbf{GPUs}: %
Although AMD and NVIDIA have recently released newer GPUs than the ones available in the systems we study, we chose to study MI60's, RTX 5000's, and V100's because they are widely used in modern HPC systems.
Prior work recommended pinning frequency, power limits, fan speed, and voltage ID to ensure that all GPUs have the same initial state (from the software and user perspective)~\cite{Scogland2014-MeasurementLevels}. 
However, Volta V100 GPUs do not have fans~\cite{cloudlab-fans}.
Instead, like many server-class GPUs, they are attached to large heatsinks and rely on server chassis fans to pull air across them for cooling.
As a result, it is not possible to adjust their fan speed (e.g., using nvidia-smi) like desktop-class GPUs.
Moreover, as we do not have administrative privileges on the clusters, we could not pin the frequency, power limit, or voltage ID.
Instead, we verified that all GPUs were configured to the maximum frequency and power limit: 1530MHz and 300W for the V100s and 1800MHz and 300W for the MI60s, respectively.
We believe this setup is still relevant because it demonstrates how much variability a common HPC user, who does not have these privileges, would see.
Moreover, we observed similar variability on a smaller CloudLab cluster where we had administrator privileges and could pin frequency and power (Section~\ref{subsec:powerlimit}).

\noindent
\textbf{Workloads}: We selected four applications from parts of CORAL-2~\cite{coral2}, a collection of benchmarks used to measure exascale workload performance.
To ensure that our workloads stress different GPU components and thus provide a holistic view of variability, we chose applications that are compute-bound, memory-bound, or balanced~\cite{AmdApuPaper, CoplinBurtscher2016-gpgpuPower, OresteVilla}.
Table~\ref{tab:summary-benchmarks} summarizes the applications.
SGEMM is a compute-intensive matrix-multiply kernel that is widely used in a number of workloads, including ML.
ResNet-50 is a popular, multi-GPU, compute-intensive ML training workload.
BERT is a popular, multi-GPU Transformer-based model that is part of the of the widely used MLPerf training benchmark suite.
LAMMPS is a popular scalable scientific computing application that is memory-bound our selected configuration.
Finally, we tested another memory-bound workload, PageRank, which is used in the Havoq graph analytics benchmark.
To ensure we study both single GPU and multi-GPU runs, we run LAMMPS, PageRank, and SGEMM on a single GPU while ResNet and BERT are run across multiple GPUs.  We also analyzed ResNet when running as a single GPU application with batch size scaled down appropriately to enable comparisons with the multi-GPU ResNet experiments.
To evaluate variability across clusters, we ran SGEMM on all clusters; we ran all other workloads on TACC's Longhorn cluster.
We provide further details about each application's setup in their corresponding Results \& Analysis sections.
Collectively, analyzing these workloads helps us gauge whether GPU variability is application-specific or not in HPC systems.

\begin{table}[tb!]
    {\footnotesize
  \centering
  \begin{tabular}{|c|c|c|c|}
    \hline
    \multirow{2}{*}{\textbf{Benchmark}} & \multirow{2}{*}{\textbf{Input Size}} & \textbf{Clusters} & \textbf{Collection} \\
    & & \textbf{Observed} & \textbf{Duration} \\ \hline
    \multirow{3}{*}{\textbf{SGEMM}\cite{cublas}} & $25536\times25536$ & Longhorn & 6 Weeks \\ %
    & $25536\times25536$ & Summit & 8 Weeks  \\ %
    & $24576\times24576$ & Corona & 2 Weeks  \\ \hline
    \multirow{2}{*}{\textbf{ResNet-50}\cite{ResNet-pyTorchRef}} & Train. Set: 1.2M images & \multirow{2}{*}{Longhorn} & \multirow{2}{*}{2 Weeks}  \\
    & Batch size: 64  & & \\ \hline
    \multirow{2}{*}{\textbf{BERT}\cite{DevlinChang18-bert}} & Train. Set: 30K words & \multirow{2}{*}{Longhorn} & \multirow{2}{*}{1 Week}  \\
    & Batch size: 64  & & \\ \hline
    \textbf{LAMMPS}\cite{LAMMPS} & $(x,y,z) = (8,16,16)$ & Longhorn & 2 Weeks  \\ \hline
    \textbf{PageRank}\cite{CheBeckmann2013-pannotia} & $643994\times643994$ & Longhorn & 2 Weeks \\ \hline
  \end{tabular}
  \vspace{-1ex}
  \caption{Summary of applications studied on HPC clusters.}
  \label{tab:summary-benchmarks}
  \vspace{-4ex}
}

\end{table}

\noindent
\textbf{Measurement}: We collected four metrics for the duration of each application:
kernel runtime (iteration runtime for ResNet-50 and BERT) in milliseconds (ms), GPU CU/SM temperature (\degree C), GPU CU/SM power consumption (Watts), and GPU CU/SM frequency (MHz).
To collect this data we use the GPU vendor's profilers~\cite{nvprof, rocprof}. %
As in prior work~\cite{Scogland2014-MeasurementLevels}, we computed the recommended sample size (number of GPUs) for each cluster to obtain $\lambda = 0.5\%$ accuracy for average power within a 95$\%$ confidence interval.
Given that we sample measurements from almost all GPUs in each cluster, our sample size is 2.9$\times$ larger than the worst-case recommendations.
Thus, our measurements are statistically significant.
Unless otherwise specified, we use the median of each measurement to avoid one-off outliers.
Since 1ms is the minimum sampling interval for these profilers, we configured our input sizes to ensure kernel durations were larger than 1ms. %
To characterize applications as compute- or memory-intensive, we also collected profiler metrics related to functional unit (FU) utilization (the utilization level of arithmetic functional units measured by nvprof on a scale of 0 to 10), DRAM utilization, and stalls.
As all the clusters we study are shared, we collected measurements while other machines were in use.
However, for consistent measurement we ensured there was no timesharing of our allocated nodes or GPUs during data collection.
By using exclusive allocations and staggered run times, we eliminated spatial and temporal effects on variability for all applications.
We discuss spatial effects further in Section~\ref{sec:takeaways}.
Finally, to protect against transient effects we collected data for multiple runs on the same machine over multiple days or weeks.

\noindent
\textbf{IQR \& Variability}: %
We use box and whiskers plots and inter-quartile
regression (IQR) to help determine variation and categorize %
statistically significant outliers.
Each box plot represents the spread from quartile 1 to 3 ($Q1$ to $Q3$).
The center of the box represents the median ($Q2$).
With $IQR$ = $Q3 - Q1$, the upper and lower whiskers in the plot represent $Q3 + 1.5 IQR$ and $Q1 - 1.5 IQR$, respectively.
The IQR captures 99.3\% of the Gaussian distribution within the box-and-whisker format.
We define \emph{range} as the difference between the upper and lower whiskers.
We denote the variation of a metric as $\frac{range}{Q2}$ and \textbf{denote all data
points that fall outside the whiskers as \emph{outliers}. Thus, outliers are \emph{not} included in our variance calculations}.

\vspace{-1ex}
\section{Variation across clusters}
\label{sec:res}

\vspace{-1ex}
\subsection{Methodology for SGEMM Application}
\label{sec:res-sgemm-methodology}

\begin{figure*}[bt!]
  \centering
  \includegraphics[width=\textwidth]{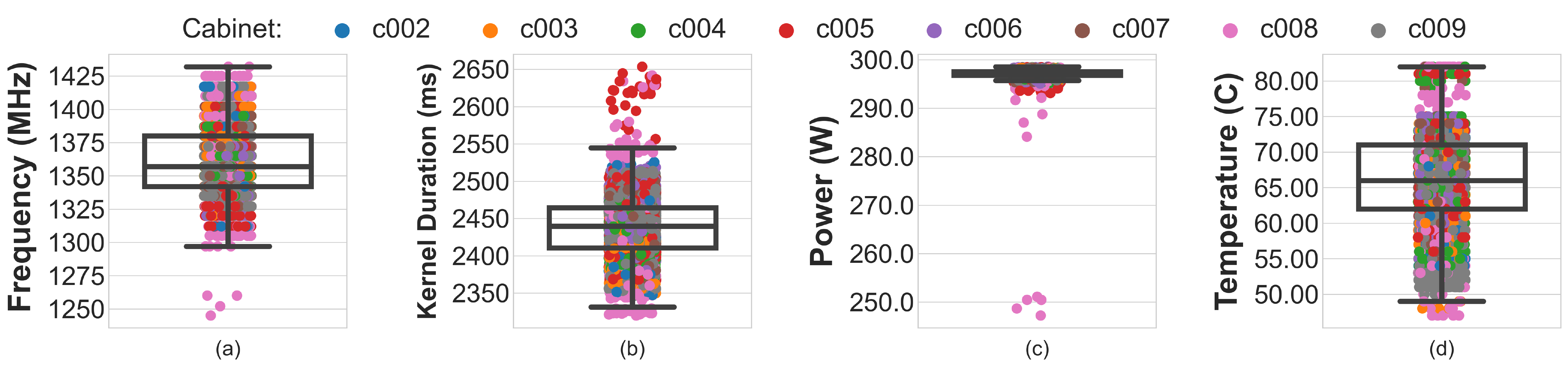}
  \vspace{-0.6cm}
  \caption{Summary results for SGEMM on Longhorn cluster presented as box plots of (a) frequency, (b) kernel duration (performance), (c) power, and (d) temperature. The color indicates the cabinet of the GPU.}
  \label{fig:tacc-summary}
  \vspace{-3ex}
\end{figure*}

\begin{figure*}[bt!]
  \centering
  \includegraphics[width=\textwidth]{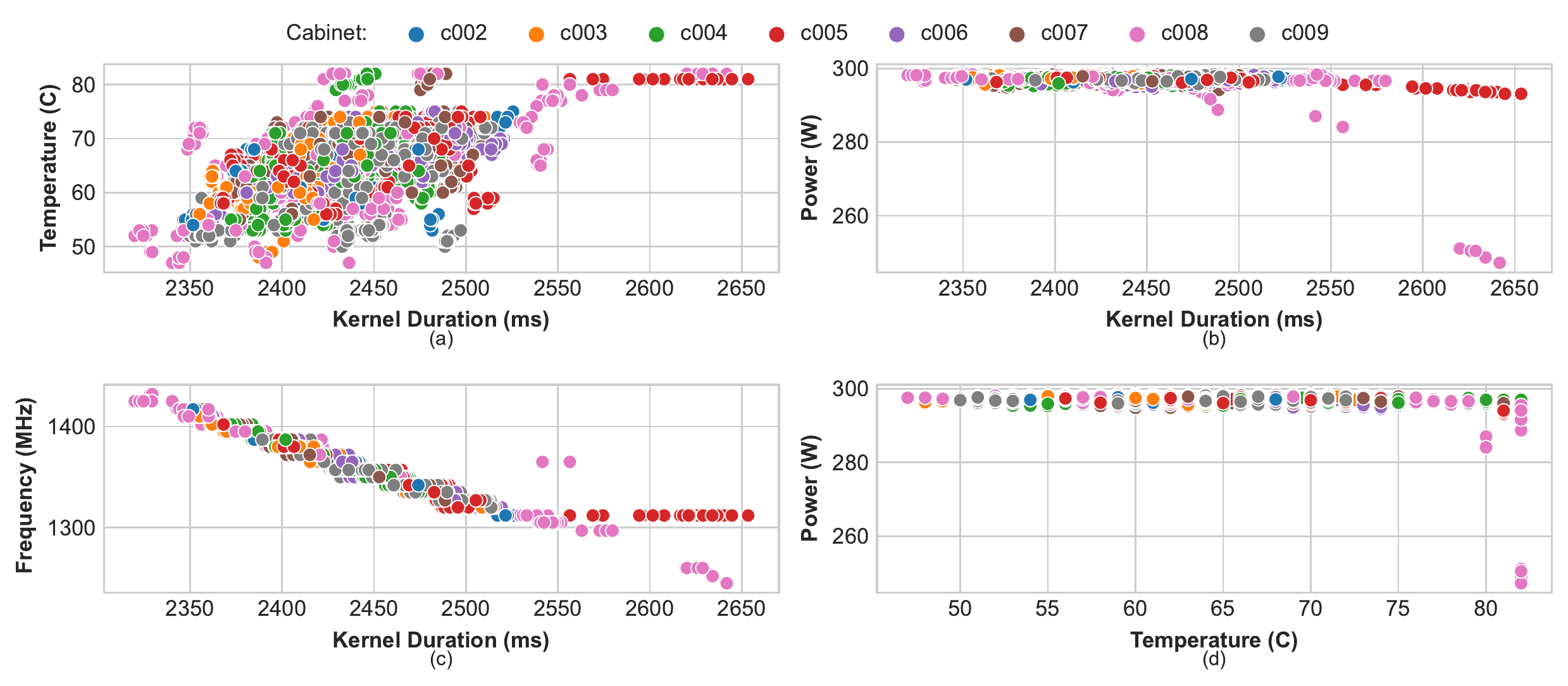}
  \vspace{-0.6cm}
  \caption{Scatter plots for SGEMM on Longhorn showing that (a) performance and temperature have a weak positive correlation (Pearson correlation coefficient $\rho = $0.46), (b) power and performance have a weak inverse correlation ($\rho = -$0.35), (c) performance and frequency are strongly correlated ($\rho = -$0.97) and (d) power and temperature are uncorrelated ($\rho = -$0.1). The color indicates the cabinet of the GPU.}
  \label{fig:tacc-correl}
  \vspace{-3ex}
\end{figure*}

We use SGEMM to study variability across clusters.
Our SGEMM application is a single SGEMM kernel that performs matrix multiplication on two matrices containing single-precision floats.
To do this, we use optimized SGEMM implementations from NVIDIA’s cuBLAS~\cite{cublas} and AMD’s hipBLAS~\cite{hipblas} libraries for the respective GPUs.
To study the effects of the GPU’s PM controller, it is important that all streaming multiprocessors (SMs) or compute units (CUs) are fully occupied and the work is evenly distributed across SMs/CUs.
Moreover, because the PM controller relies on DVFS to maintain the power limit for safe operation~\cite{GeVogt2013-dvfsKepler, v100}, the kernel run must be long enough for the DVFS controller to reach a stable state~\cite{CoplinBurtscher2016-gpgpuPower}.
Thus, we carefully tuned the matrix size (Table~\ref{tab:summary-benchmarks}) for both NVIDIA V100s and AMD MI60s to (i) achieve a sufficient runtime (Section~\ref{sec:method}), (ii) ensure high performance, and (iii) provide high SM/CU occupancy.

We define 1 run of our experiment as 100 repetitions of the SGEMM kernel.
The repetitions help avoid statistical bias and transient effects.
Before collecting data, we run one warm-up run to avoid counting cuDNN startup overheads~\cite{ChouNg2020-dab, cudnn-caching}.
We provide the exact same matrix inputs to every GPU, since the data inputs themselves are not important in our study.%

\vspace{-1ex}
\subsection{SGEMM on TACC Longhorn}
\label{sec:res-longhorn}

\begin{figure*}[tb!]
  \centering
  \begin{subfigure}{.4\textwidth}
    \centering
    \includegraphics[width=\textwidth]{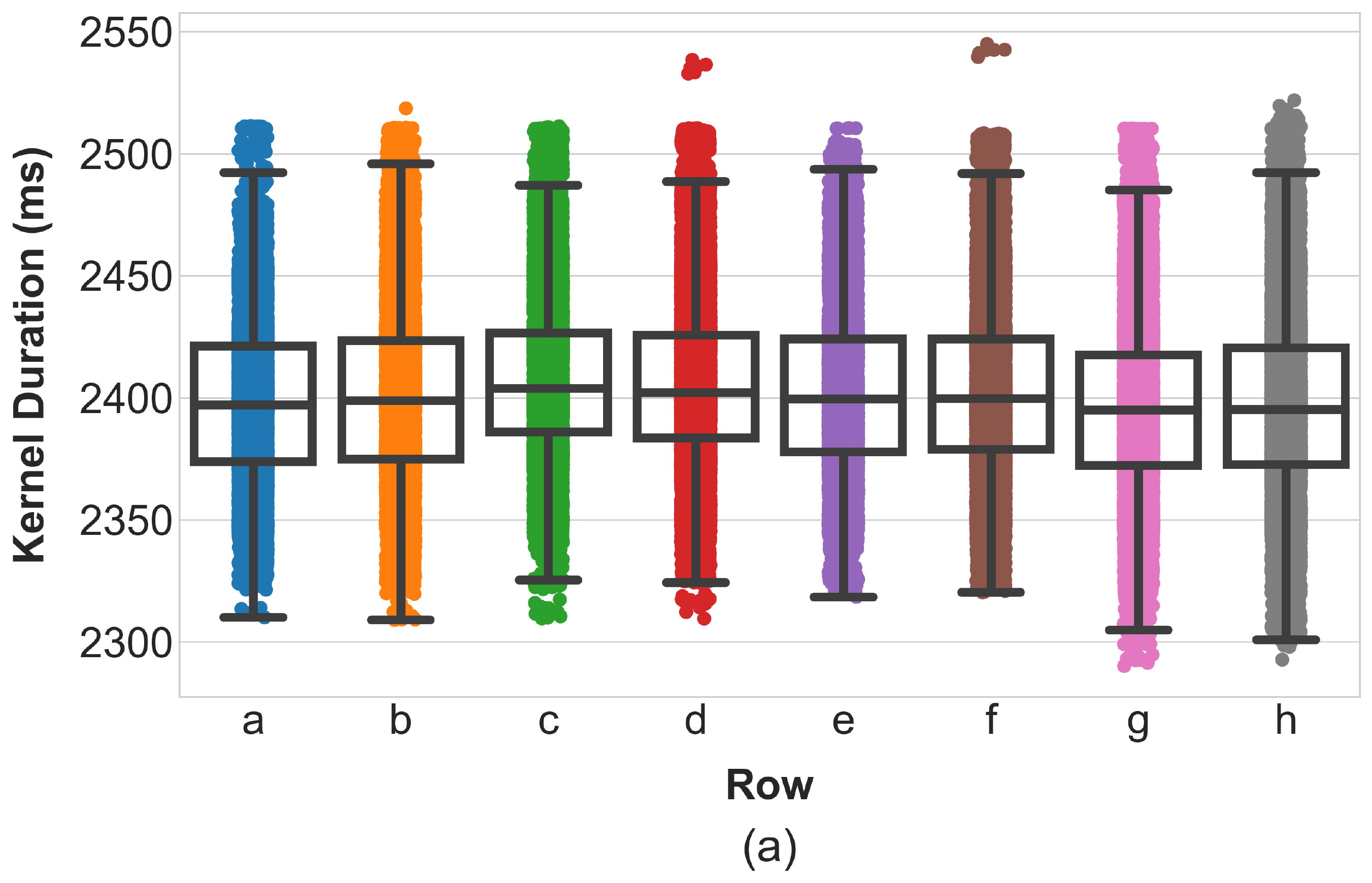}
  \end{subfigure}
  \begin{subfigure}{.4\textwidth}
    \centering
    \includegraphics[width=\textwidth]{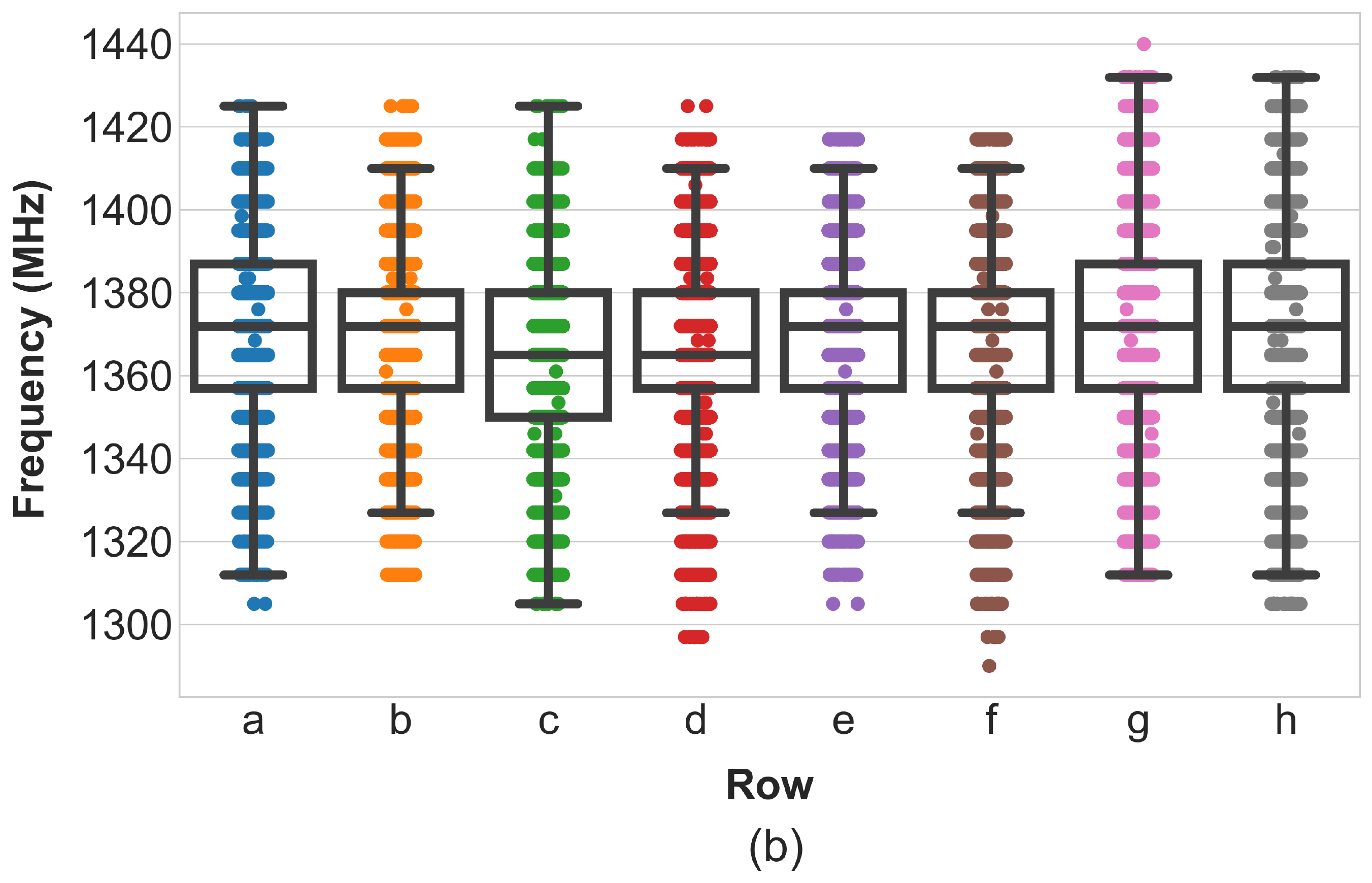}
  \end{subfigure}
  \begin{subfigure}{.4\textwidth}
    \centering
    \includegraphics[width=\textwidth]{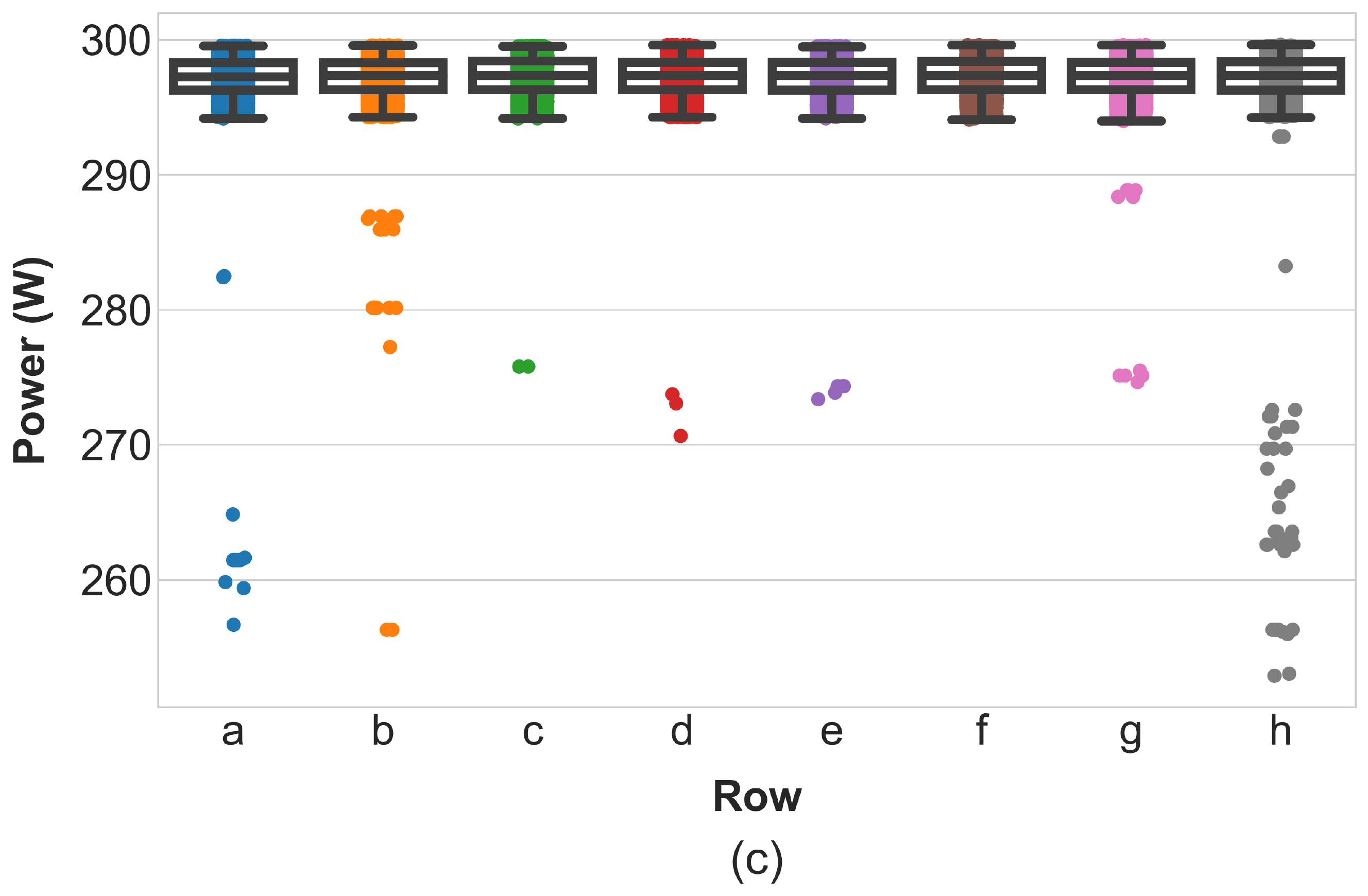}
  \end{subfigure}
  \begin{subfigure}{.4\textwidth}
    \centering
    \includegraphics[width=\textwidth]{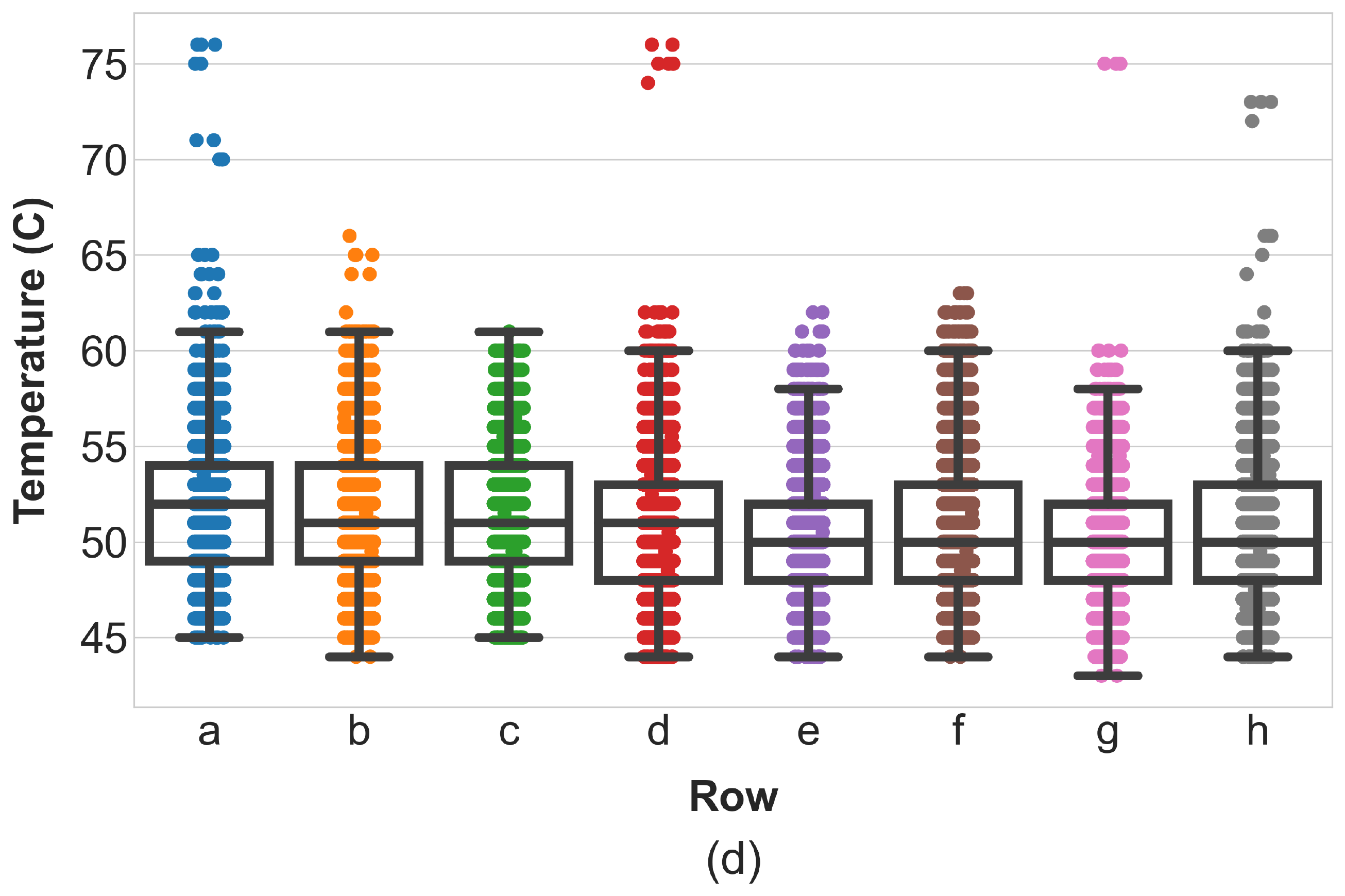}
  \end{subfigure}
  \vspace{-0.3cm}
  \caption{Summary results for SGEMM on Summit, showing variation in (a) kernel duration (performance), (b) frequency, (c) power, and (d) temperature. We breakdown box plots by row because of the large scale of the cluster.}
  \label{fig:summit-row-summary}
  \vspace{-3ex}
\end{figure*}

Figure~\ref{fig:tacc-summary} shows aggregated box plots for kernel duration, frequency, power, and temperature for SGEMM on Longhorn.
Overall, SGEMM has 9\% performance variation on Longhorn. %
Moreover, despite verifying that the GPUs are configured to run at the highest frequency (1530MHz), Figure~\ref{fig:tacc-summary} shows that the GPUs run at lower frequencies (1300-1440MHz).
The overall frequency variation is 140MHz or 11\%.
We also observe a wide spread in temperatures (33\degree C between $Q1$ and $Q3$).
Furthermore, certain GPUs also run at power levels far below 300W: at around 250W.
This significant variation in performance, power, and temperature motivated us to further investigate the relationship between these metrics.

Figure~\ref{fig:tacc-correl} presents scatter plots between the distinct measurement pairs.
In Figure~\ref{fig:tacc-correl}, although sometimes the slowest kernels run on the GPUs at the highest temperatures -- and the fastest kernels run on the lowest temperature GPUs -- overall performance and temperature are not strongly correlated ($\rho =$ 0.46).
For example, multiple GPUs running at the same temperature have up to a 200ms (10\%) performance difference.
Additionally, the top-right of the performance vs temperature plot shows a cluster of GPUs with both high temperature and longer runtimes.
This implies that if a GPU is running at a high temperature ($>$78\degree C), it is likely to have worse performance.
However, this is not always the case, because other GPUs run at similar temperatures but complete much faster (e.g., c004).
This is surprising because as frequency increases, temperature is expected to increase.
Moreover, usually higher frequencies improve SGEMM's performance~\cite{GeVogt2013-dvfsKepler, Scogland2015-pwrPerspectives} and our results agree with this trend. Figure~\ref{fig:tacc-correl}c shows that performance and frequency are strongly correlated on Longhorn ($\rho = -$0.97).
Further, Figure~\ref{fig:tacc-correl}b shows that certain GPUs consume less than 290W, and these GPUs usually have lower performance and higher temperatures.
This indicates that higher temperature levels may hurt performance.
OVerall, these variations in trends highlighted the need to examine these GPUs more closely.
So, we reported our findings to Longhorn system administrators, which helped them perform early identification of a poorly performing node and investigate it in greater detail.

\begin{figure*}[tb!]
  \centering
  \includegraphics[width=\textwidth]{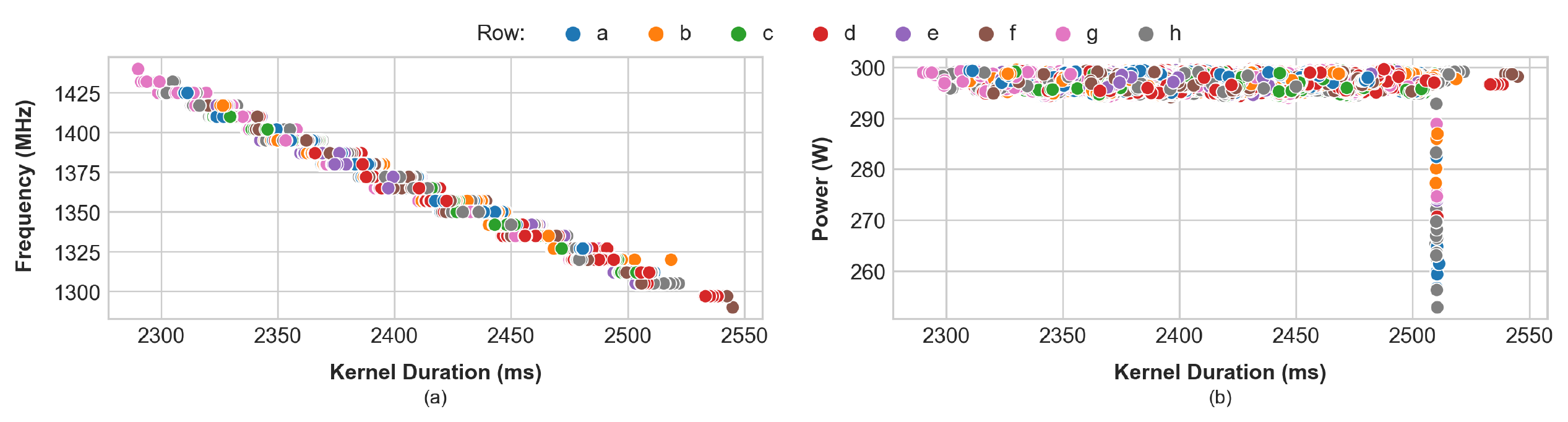}
  \vspace{-0.6cm}
  \caption{Scatter plots of Summit data showing (a) performance and frequency have strong negative correlation ($\rho = -$0.99) and (b) performance and power have almost no correlation ($\rho = -$0.09). The color indicates the row for the GPU.}
  \label{fig:summit-scatterplots}
  \vspace{-3ex}
\end{figure*}

\noindent\textbf{Takeaway 1}: \emph{Despite running the same kernel on similarly configured GPUs,
we see 9\% performance variance across GPUs in the Longhorn cluster, with temperature range
$\geq$ 30$\degree C$ and some power outliers at 250W.} 

\subsection{SGEMM on ORNL Summit}
\label{subsec:res-summit}

We next examine the larger Summit to see if similar patterns are seen. As shown in Table~\ref{tab:summary-hpc-clusters}, Summit %
contains 27648 GPUs and is water cooled.
We collect the same measurements for SGEMM as Longhorn.
However, since Summit has so many GPUs, we further breakdown our measurements based on the particular row and column where a machine resides (using node location references derived from ORNL's layout~\cite{summit-layout}).

Figure~\ref{fig:summit-row-summary} shows aggregated kernel duration, frequency, power, and temperature box plots, grouped by rows.
Similar to Longhorn, Summit has 8\% performance variation across all rows, with rows D and F having the most outliers (Figure~\ref{fig:summit-row-summary}a).
Likewise, frequency variation (Figure~\ref{fig:summit-row-summary}b) is again near 100MHz across all rows, although rows D and F have some outliers %
below 1300MHz.
Moreover, although all row's IQRs (Figure~\ref{fig:summit-row-summary}c) range from 295-300W, a number of GPUs consume less than 290W, especially in rows A and H.
However, unlike Longhorn, Summit's temperature range is narrow: 40\degree-62\degree C.
This shows the benefits of water cooling~\cite{ostrouchov2020gpulifetimes} compared to Longhorn, although water cooling does not significantly impact frequency or performance variation.

Figure~\ref{fig:summit-scatterplots} shows Summit's scatter plots. %
Similar to \textbf{Takeaway 1}, performance and frequency were directly correlated ($\rho = -$0.99) on Summit (Figure~\ref{fig:summit-scatterplots}a).
However, unlike Longhorn, Summit has a string of power outliers below 290W with 2510ms runtime (Figure~\ref{fig:summit-scatterplots}b).
This is unexpected, since GPUs with a wide power range typically vary in their performance.

To understand these outliers, we analyzed row H, column 36 because of its variance in power consumption.\footnote{See Appendix~\ref{sec:apdx-ornl} graphs for this row-column pair.}
We found that 7 of its nodes %
exhibited power outliers, with power as low as 255W, while the remaining 9 nodes did not have any outliers.
However, despite the power outliers, nodes 10 and 11 did not have any temperature outliers, unlike Longhorn.
Hence, these two nodes show that although water cooling decreases temperature variation, it does not prevent nodes from having large performance and power variation.
Moreover, the variance in power consumption seen in row H, column 36 highlights the difficulty in drawing conclusions from cluster-wide summaries and provides another example of how our study can be used to flag underperforming nodes early for system administrators to investigate further, similar to Longhorn.

\noindent\textbf{Takeaway 2}: \emph{Summit and Longhorn have similar performance and frequency variation trends, but Summit has severe power outliers that are concentrated in a few rows.}

\noindent\textbf{Takeaway 3}: \emph{While Summit and Longhorn have the same GPU temperature setups, water cooling reduces temperature variation, but does not improve performance and power variation.}

\subsection{SGEMM on Corona LLNL}
\label{sec:res-sgemm-corona}

\begin{figure*}[tb!]
    \centering
    \includegraphics[width=\textwidth]{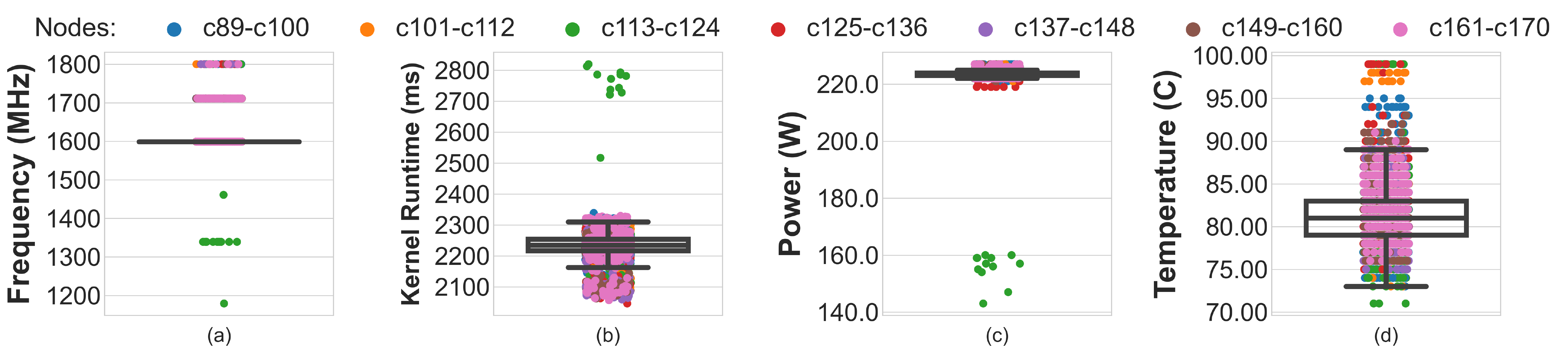}
    \vspace{-0.6cm}
    \caption{Summary results for SGEMM on Corona presented as box plots of (a) frequency, (b) mean kernel duration (performance), (c) power, and (d) temperature. Node c115 (green) is the outlier.}
    \label{fig:sgemm-corona-summary}
    \vspace{-2ex}
\end{figure*}

\begin{figure*}[tb!]
    \centering
    \includegraphics[width=\textwidth]{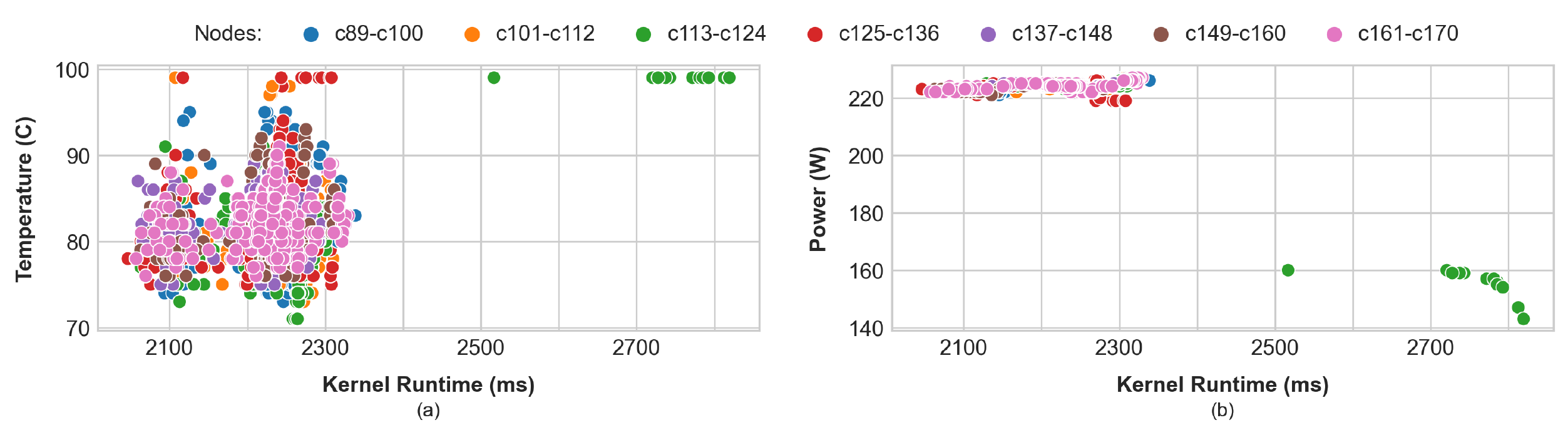}
    \vspace{-0.6cm}
    \caption{Scatter plots for SGEMM on Corona showing correlations
    between (a) kernel duration (performance) and temperature  
    ($\rho = $0.20) and (b) kernel duration and power
    ($\rho = -$0.48). Node c115 (green) is the outlier.}
    \label{fig:sgemm-corona-correl}
    \vspace{-2ex}
\end{figure*}

Since compute clusters use GPUs from multiple vendors, it is important that we also examine how AMD GPUs vary.
Corona uses air-cooling for its AMD MI60 GPUs.
We collected the same measurements as Longhorn and Summit.
However, unlike Longhorn and Summit we were unable to obtain a system map and thus analyze our data by node.
Nevertheless, to improve readability we group nodes in ``cabinets" of 12 GPUs, like the similarly-sized Longhorn cluster.

Figure~\ref{fig:sgemm-corona-summary} presents aggregated box plots for our 4 metrics.
Individual data points are distinguished by our node groupings.
Overall, Corona exhibits 7\% runtime variation (Figure~\ref{fig:sgemm-corona-summary}b), similar to Longhorn (9\%) and Summit (8\%).
Interestingly, frequency and performance are not as strongly correlated on Corona ($\rho=-$0.76 on Corona, versus Longhorn's $-$0.93 and Summit's $-$0.99), as Figure~\ref{fig:sgemm-corona-summary}a shows much less variability than Longhorn and Summit (\textbf{Takeaways 1-2}).
Moreover, the MI60s have coarser frequency levels than the NVIDIA V100s, suggesting that the GPU vendor's DVFS schemes vary significantly.
Finally, the power (2\%, Figure~\ref{fig:sgemm-corona-summary}c) and temperature (20\%, Figure~\ref{fig:sgemm-corona-summary}d) variability are similar to Longhorn.
However, whereas Longhorn has more power outliers consuming as little as 250W, Corona only has one outlier node (c115) which consumes 165W.
Surprisingly Corona's nodes never reach the max power of 300W.
Since the temperature are as high as 99\degree C (near the slowdown temperature), we believe the DVFS controller heavily throttles the frequency (most frequencies also do not reach the peak 1800MHz), thereby decreasing the power since $P_{dynamic} \propto frequency$.

To further analyze the outlier node, c115, Figure~\ref{fig:sgemm-corona-correl} correlates our metrics.
Corona's kernel runtime and power relationship (Figure~\ref{fig:sgemm-corona-correl}b) is similar to Longhorn and matches expectations: slower GPUs tend to consume less power.
However, Summit does not exhibit this trend, as similarly performing GPUs in Summit exhibit a wide power range.
The performance-power correlation coefficients are $-$0.48 and $-$0.35 on the similarly-sized Corona and Longhorn clusters, respectively, but $-$0.09 on Summit.
However, since the cluster sizes differ, we also compared Summit against a scaled normal distribution of Longhorn's performance numbers to determine cluster size's impact.
The scaled normal performance distribution projects that the Longhorn data would have 9.4$\%$ variability on a Summit-sized cluster.
Since our actual Summit measurements (Section~\ref{subsec:res-summit}) had 8$\%$ performance variability, this 
suggests the cluster size may impact the severity of variability.
Also, while c115 runs very hot compared to most nodes (Figure~\ref{fig:sgemm-corona-correl}a), there is not a clear relationship between kernel runtime and temperature ($\rho = $0.20).
Typically higher temperatures occur when GPUs run at higher frequencies and achieve better performance.
However, there are several nodes that run as hot as c115 but which do not perform worse than the median GPU, which likely rules out that this outlier was due to inadequate air cooling.
Moreover, all of these GPU's temperatures are near the slowdown temperature.
Additionally, Summit nodes that perform similarly also exhibit a wide temperature range.
For example, Corona GPUs performing around 2210ms have a 25\degree C temperature range.
Thus, c115 appears to be an underperforming GPU and another candidate for further inspection and potential replacement by the system administrators.

\begin{figure}[tb!]
  \begin{centering}
  \includegraphics[width=0.8\columnwidth]{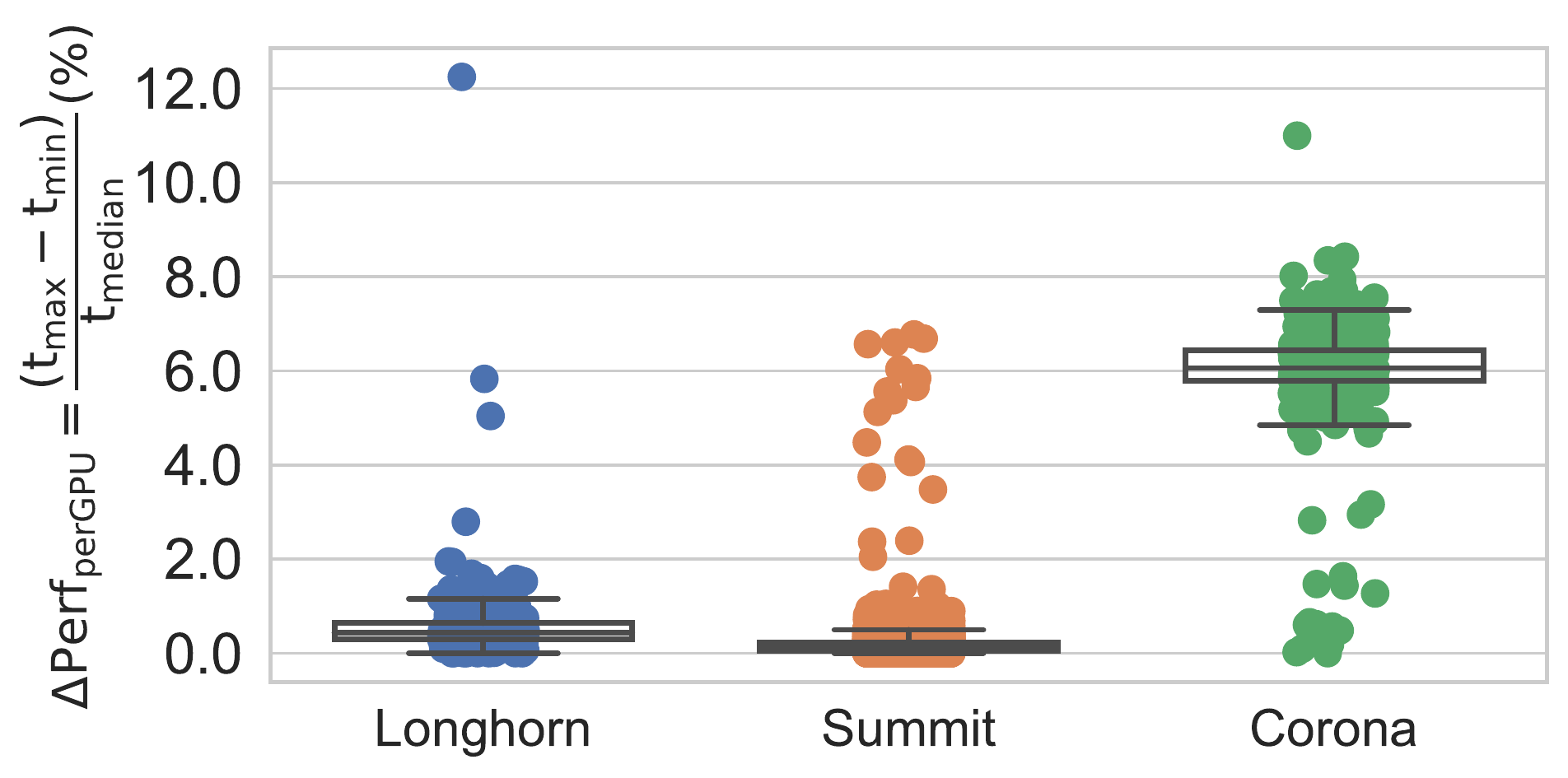}
  \caption{Normalized performance variation within a GPU for SGEMM on Longhorn, Summit, and Corona clusters.}
  \label{fig:per-gpu-var-boxplot}
  \end{centering}
  \vspace{-4ex}
\end{figure}

Lastly, a single GPU exhibits little performance variation across independent application runs. 
Figure~\ref{fig:per-gpu-var-boxplot} shows the normalized per-GPU performance variance for SGEMM on various clusters.
The median variance values are 0.44$\%$, 0.12$\%$ and 6.06$\%$ on Longhorn, Summit, and Corona, respectively.
Thus, our results are repeatable and any transient effects cause only minor variations.
Although there are a few outliers as high as 12$\%$ (Longhorn node c002-010), we found that these do not correspond to the worst performing GPUs.
Instead, they fall between the median and $Q3$, which suggests that ill-performing GPUs are consistently ill-performing.

\noindent\textbf{Takeaway 4}: \emph{Corona's AMD GPUs exhibit similar behavior to the like-sized Longhorn cluster with NVIDIA GPUs. However, the performance-power relationship in the larger Summit differs from Corona and Longhorn, suggesting that cluster size affects variability degree.}  %

\subsection{SGEMM on SNL Vortex}
\label{sec:res-vortex}

\begin{figure*}[tb!]
    \centering
    \includegraphics[width=\textwidth]{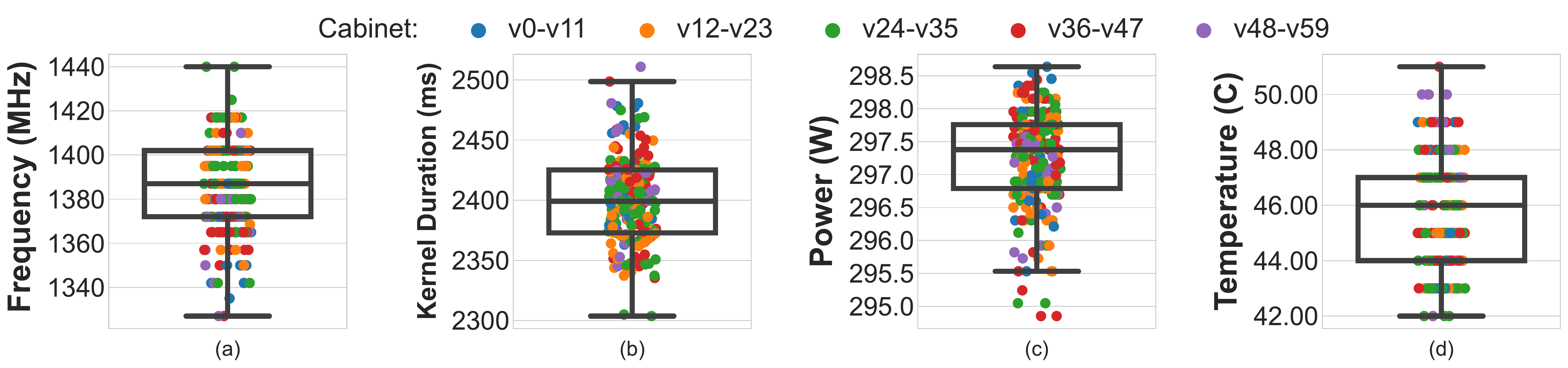}
    \vspace{-0.6cm}
    \caption{SGEMM box plot summary results on SNL Vortex for (a) frequency, (b) performance, (c) power, and (d) temperature.}
    \label{fig:snl-summary}
    \vspace{-3ex}
\end{figure*}

We also evaluated SGEMM's behavior on SNL's Vortex cluster~\cite{snl}.
Vortex is similarly sized to Longhorn and Corona, but is water-cooled whereas Longhorn and Corona are air-cooled.
We gathered data from 184 GPUs on Vortex.
However, we were unable to obtain information about their location within the cluster.
Thus, similar to Corona, we simplify readability of plots by grouping nodes into ``cabinets" of 12 GPUs.

Figure~\ref{fig:snl-summary} shows aggregated box plots for kernel duration, frequency, power, and temperature when the SGEMM kernel is run on Vortex.
We observed a performance variation (9\%) similar to Longhorn (9\%) and Summit (8\%) clusters.
Despite the GPUs being configured to run at the highest frequency level (1530MHz), Figure~\ref{fig:snl-summary} shows that they often run at lower frequencies (1330-1442 MHz), similar to other clusters.
These frequencies vary by about 100 MHz or 10\% between the fastest and slowest GPUs observed.
The temperature difference between the GPUs has a narrower spread (10\degree C between $Q1$ and $Q3$) on Vortex than Longhorn.
This is likely due to Vortex's use of water-cooling.
From Figure~\ref{fig:snl-summary}c, we also observe that all GPUs operated within 5W of the 300W power limit, unlike Longhorn where some outliers consuming significantly lower power (Figure~\ref{fig:tacc-summary}). 

\begin{figure*}[tb!]
    \centering
    \includegraphics[width=\textwidth]{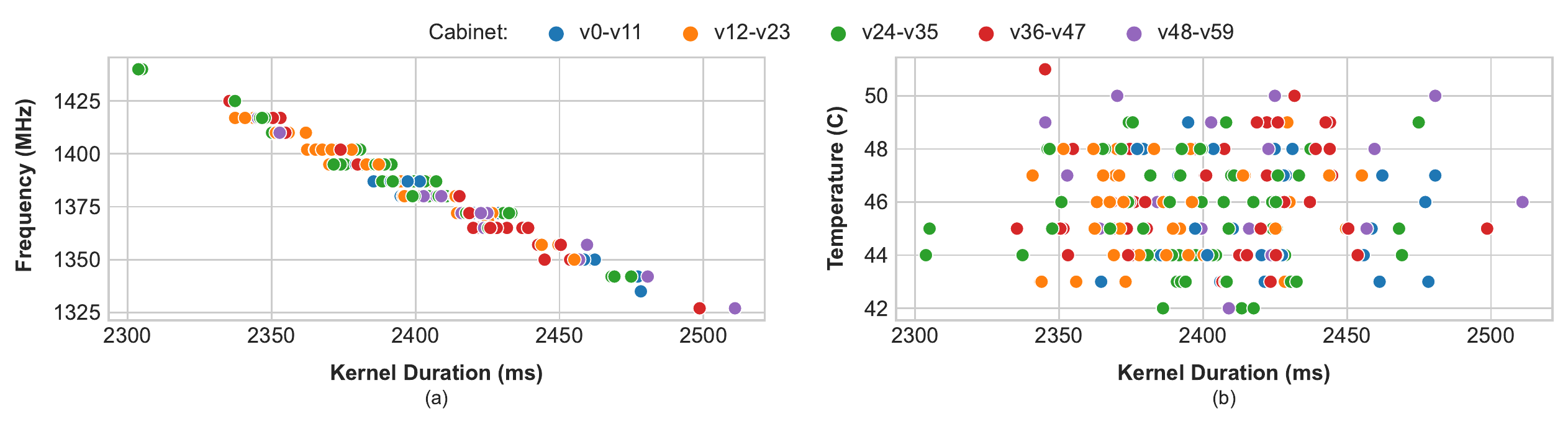}
    \vspace{-0.6cm}
    \caption{Scatter plots for SGEMM on SNL Vortex showing (a) kernel duration (performance) and frequency have a strong negative correlation ($\rho = -$0.98) 
    and (b) kernel duration and temperature are uncorrelated
    ($\rho = $0.04).}
    \label{fig:snl-correl}
    \vspace{-2ex}
\end{figure*}

Figure~\ref{fig:snl-correl} presents correlations between the metrics.
Similar to \textbf{Takeaway 1}, on Vortex we again see a strong inverse correlation between frequency and performance.
Figure~\ref{fig:snl-correl}b shows the correlation between temperature and kernel duration.
Similar to Longhorn (Figure~\ref{fig:tacc-correl}), multiple GPUs running at the same temperature have up to a 200 ms (10\%) difference in performance.
However, unlike Longhorn we do not observe GPUs with high temperature having longer runtimes in Vortex.
Thus, temperature and performance apper to be weakly correlated in water-cooled systems.
This may be because all observed GPUs are running relatively cooler (median 46\degree C) when compared to Longhorn (median 66\degree C).

\begin{figure*}[tb!]
  \centering
  \begin{subfigure}{\textwidth}
    \centering
    \includegraphics[width=\textwidth]{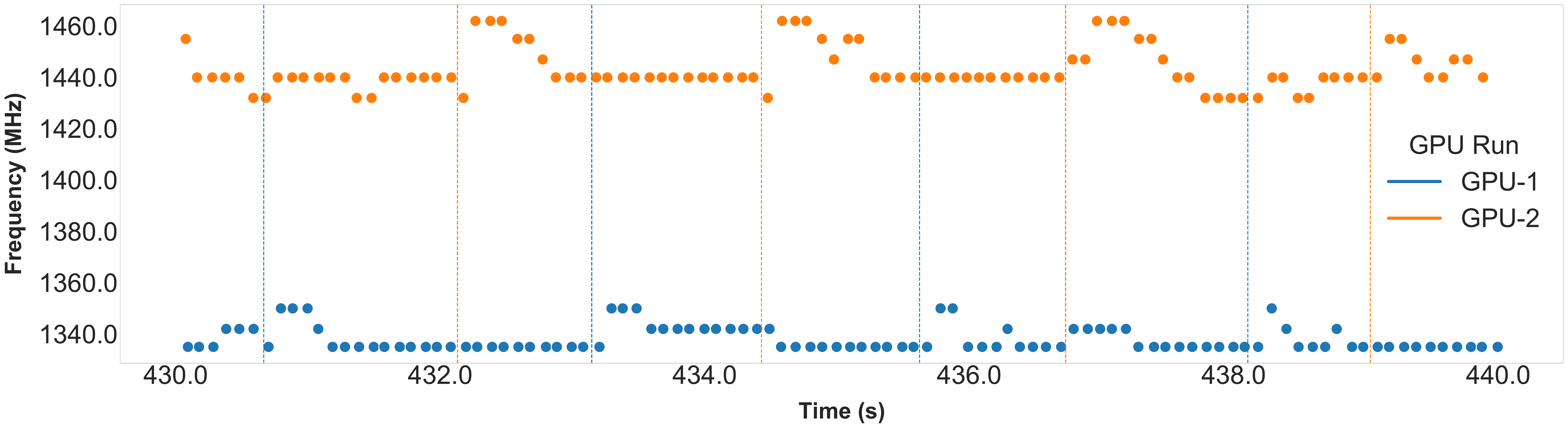}
    \caption{Frequency}
    \label{fig:snl-timeSeries-freq}
  \end{subfigure}
  \hspace{6ex}
  \begin{subfigure}{\textwidth}
    \centering
    \includegraphics[width=\textwidth]{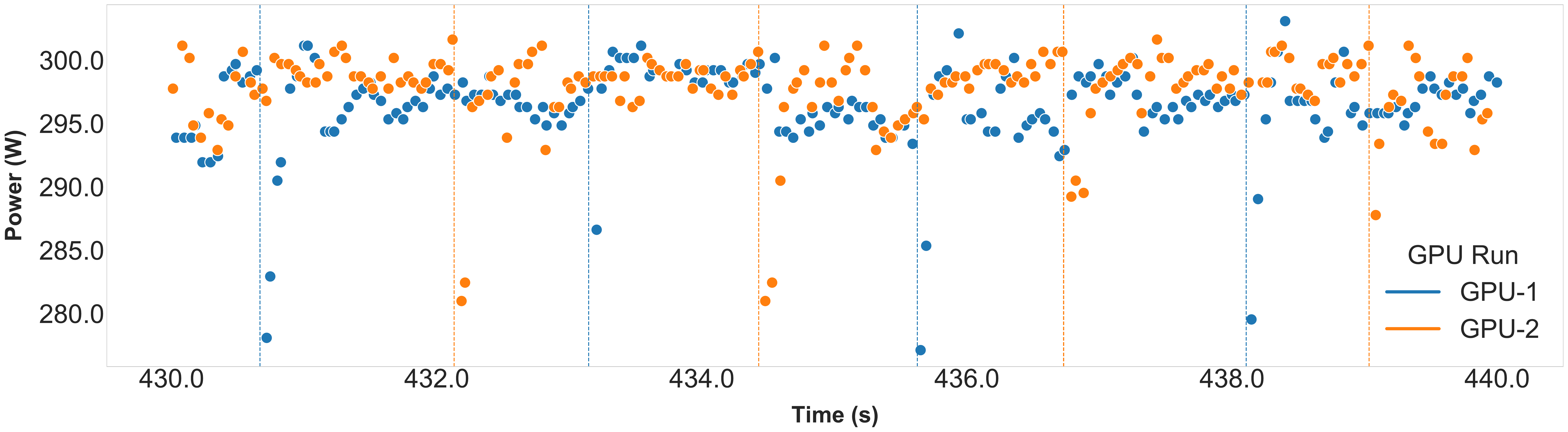}
    \caption{Power}
    \label{fig:snl-timeSeries-pwr}
  \end{subfigure}
  \vspace{-0.3cm}
  \caption{Time-series plots for (a) frequency and (b) power for two distinct GPUs executing the SGEMM Kernel on SNL Vortex. Vertical lines indicate start of a new kernel.}
  \label{fig:snl-timeSeries}
\end{figure*}

From these observations, we find a consistent trend across Longhorn, Summit, Corona, and Vortex of performance variations being inversely correlated with changes in frequency.
To verify that these frequency changes occur due to DVFS mechanisms employed by PM infrastructure, we examine the timelines of frequency and power for two GPUs at the extremes of kernel performance (Figure~\ref{fig:snl-timeSeries}).
We examine a 10s slice in the middle of the overall application run, during which 4 SGEMM kernels are launched and run to completion.
Vertical lines indicate the start time of a new kernel. Figure~\ref{fig:snl-timeSeries-freq} shows that an SGEMM kernel is launched at 430.6s and the GPU reports a frequency of 1340MHz.
Similarly, a kernel is launched on GPU-2 at 432s, with the GPU reporting an initial frequency slightly under 1440MHz.
As the kernels begin executing, the frequency starts to rise along with an increase in power draw (Figure~\ref{fig:snl-timeSeries-pwr}).
However, as soon as power reaches and exceeds the 300W TDP, the DVFS mechanism triggers a frequency drop until the GPU reaches a state where the power draw remains below 300W.
Note that DVFS affects both GPUs but GPU-1 crosses the power limit at much lower frequencies than GPU-2 (GPU-1 runs at a median frequency of 1327 MHz, whereas GPU-2 is running at 1440 MHz).
This difference in frequency translates into a 173 ms or 8\% difference in performance between the two GPUs, even when the GPUs have the same temperature (44\degree C on average) and nearly the same average power draw. 
These timelines demonstrate that (1) differences in performance between GPUs is correlated to changes in frequency, and that (2) these frequency changes are driven by the PM infrastructure trying to keep power consumption under the GPU's TDP.

Overall, the findings on Vortex also reinforce \textbf{Takeaway 3} -- water cooling reduces temperature variation but does not diminish performance or power variability.

\subsection{SGEMM on TACC Frontera}
\label{sec:res-frontera}

Finally, we study TACC's Frontera cluster.
As specified in Table~\ref{tab:summary-hpc-clusters}, Frontera is a mineral oil-cooled system with Turing-class RTX 5000 GPUs~\cite{tacc, rtx5000}.
We compare results on Frontera with that of the similarly sized Longhorn (air-cooled V100 GPUs) and Vortex (water-cooled V100 GPUs) clusters.

\begin{figure*}[tb!]
    \centering
    \includegraphics[width=\textwidth]{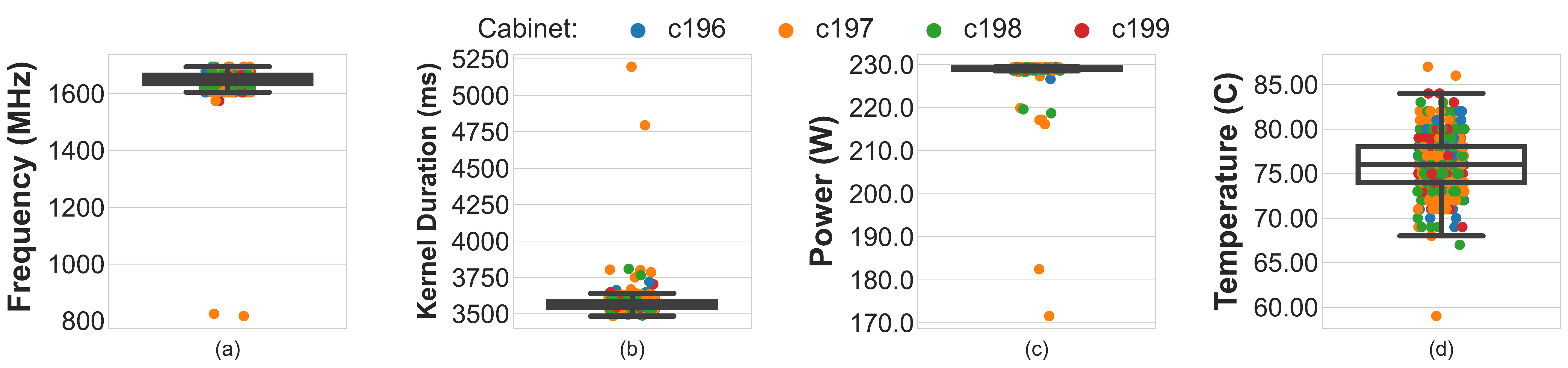}
    \vspace{-0.6cm}
    \caption{SGEMM box plot summary results on Frontera for (a) frequency, (b) performance, (c) power, and (d) temperature.}
    \label{fig:frontera-summary}
    \vspace{-3ex}
\end{figure*}

Figure~\ref{fig:frontera-summary} shows aggregated box plots for our metrics, grouped by cabinet number.
Overall, Frontera shows 5$\%$ performance variation and 7$\%$ frequency variation.
Since Quadro RTX GPUs have a faster boost clock~\cite{rtx5000} than Volta V100s, Frontera's range of operating frequencies is higher than other clusters.
Turing-class GPUs also have a lower TDP (230W~\cite{rtx5000}), and almost all GPUs on Frontera operate within 5W of this limit, similar to Vortex and Longhorn.
However, two GPUs in cabinet c197 are outliers -- they run 1100-1600ms slower than the median kernel duration, 16\degree C cooler than the median temperature, and consume 59W lesser than the median power consumption.
These outliers are similar to those on Longhorn, Summit, and Corona.
Moreover, similar to Corona, presenting this data to system administrators led them to flag the the pump that stirs the mineral oil in this cabinet for additional investigation -- further demonstrating the benefits of our approach in helping operators manage cluster operation (discussed further in Section~\ref{sec:takeaways}).
Similar to the water-cooled Vortex, Frontera has a narrow temperature spread ($Q3 - Q1 =$ 4\degree C), but the median temperature (76\degree C) is much higher than Vortex (46\degree C).
These temperature results indicate that mineral-oil cooling lies somewhere between air and water-cooling in effectiveness at reducing thermal variation.

\begin{figure*}[tb!]
    \centering
    \includegraphics[width=\textwidth]{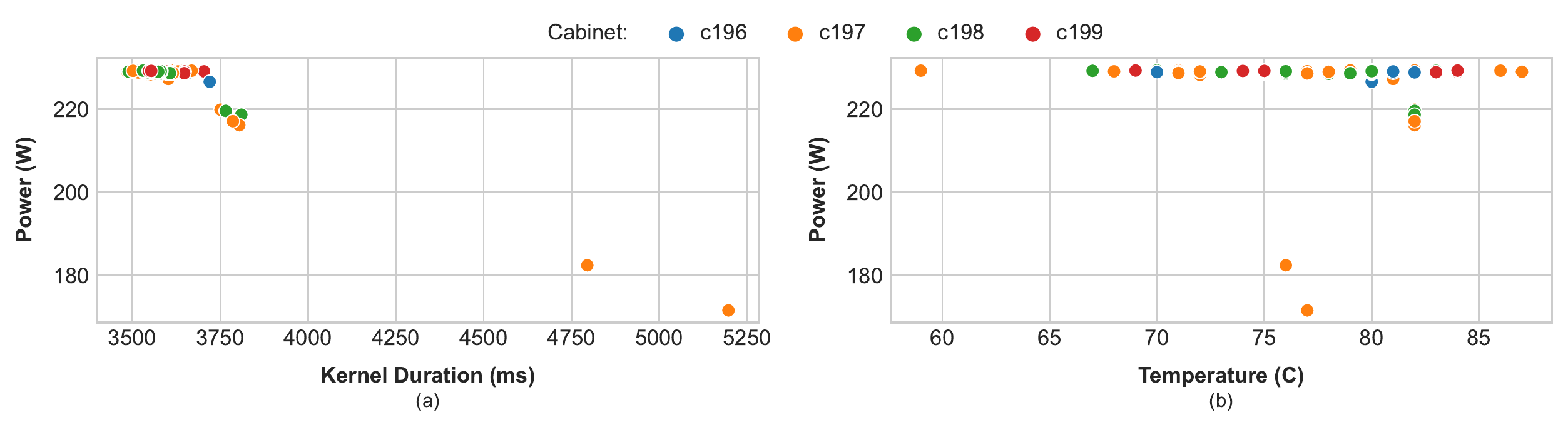}
    \vspace{-0.6cm}
    \caption{Scatter plots for SGEMM on Frontera showing that (a) kernel duration (performance) and power have a strong negative correlation ($\rho = -$0.96) 
    and (b) power and temperature are almost uncorrelated ($\rho = -$0.1)}
    \label{fig:frontera-correl}
    \vspace{-2ex}
\end{figure*}

Finally, Figure~\ref{fig:frontera-correl} presents scatterplots that correlate our metrics.
Despite the c197 outliers, there is a strong correlation between performance and power/frequency ($\rho = -$0.96).
However, Figure~\ref{fig:frontera-correl}b shows that GPUs consuming similar amounts of power can have widely varying temperatures ($\rho = -$0.1).
This plot, together with observations from Vortex, suggests that temperature correlation with other metrics is weaker in systems cooled by water or mineral oil, as opposed to air-cooled clusters.
Finally, the outlier on Frontera is concentrated to one cabinet and performs significantly worse, similar to outliers we observed on Summit (\textbf{Takeaway 2}).

\section{Variation across applications}
\label{sec:res-vary}

Overall, our study with SGEMM demonstrated that we can see significant variation across different scales, GPU vendors, and cooling approaches.
However, SGEMM is a single computationally intensive kernel.
Thus, although SGEMM is widely used in modern GPU applications, it may not be representative of larger applications.
Accordingly, we next study different applications to determine whether variability is application-specific or not.

\subsection{ResNet-50 on TACC Longhorn}
\label{sec:res-resnet-longhorn}

\begin{figure*}[tb!]
    \centering
    \includegraphics[width=\textwidth]{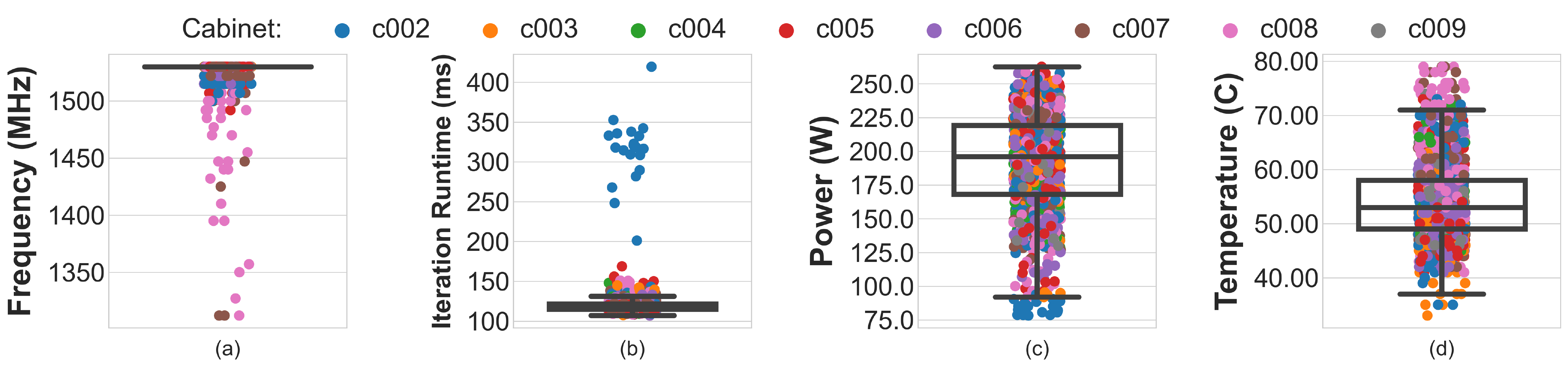}
    \vspace{-0.6cm}
    \caption{Multi-GPU ResNet-50 Longhorn box plot summaries for (a) frequency, (b) performance, (c) power, and (d) temperature.}
    \label{fig:resnet-tacc-summary}
    \vspace{-2ex}
\end{figure*}

\begin{figure*}[bt!]
    \centering
    \includegraphics[width=\textwidth]{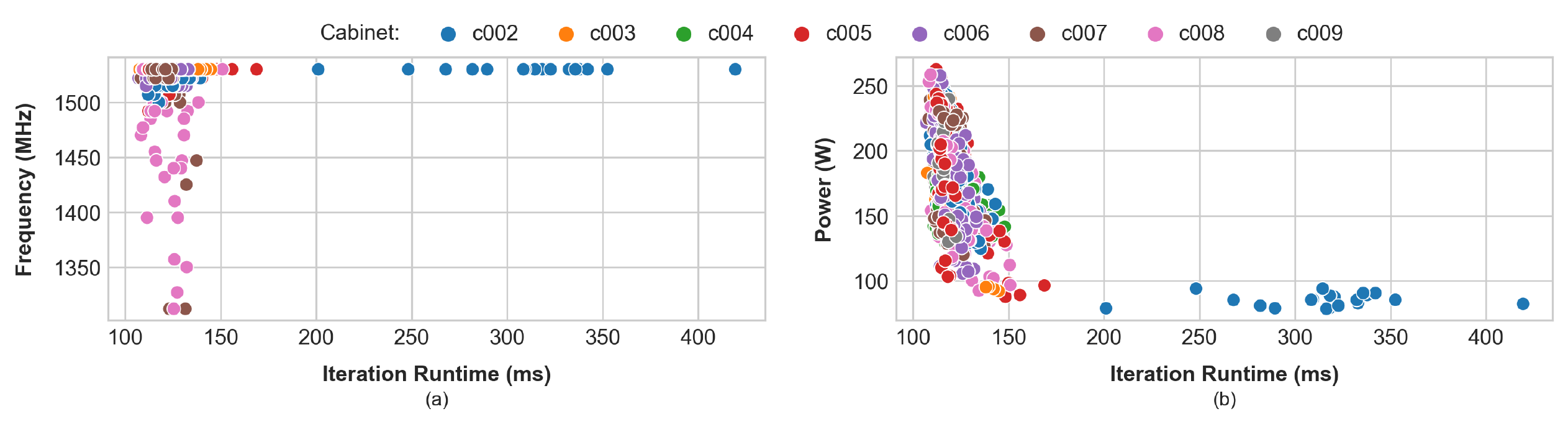}
    \vspace{-0.6cm}
    \caption{Scatter plots for multi-GPU ResNet-50 experiments on Longhorn showing (a) iteration duration (performance) and frequency are almost uncorrelated ($\rho=-$0.01) and (b) iteration duration and power have a negative correlation ($\rho = -$0.48).}
    \label{fig:resnet-tacc-correl}
    \vspace{-3ex}
\end{figure*}

Given their widespread use on GPU-rich clusters, it is important to understand performance variability for ML applications.
Thus, we studied the ResNet-50 CNN for the most intensive training phase.
We chose the 50-layer version because it is a stable, commonly used benchmark in the HPC community~\cite{MattsonReddi2020-mlPerf, MattsonCheng2020-mlPerfTraining}.
In general, ResNet-50 is computationally intensive, especially for its kernels that perform convolution.
However, ResNet-50 also performs other, more memory intensive operations, which reduce its overall computational intensity compared to SGEMM.
This is confirmed by examining FU utilization: averaging across kernels, ResNet-50's was 5.4 (on a 1$-$10 scale) while SGEMM's FU utilization was 10.  This also aligns with observations from prior work on classifying GPU applications~\cite{Guerreiro-appClasses}.
Our training set uses 1.2 million images from ImageNet with batch size 64.
We define one training run as 500 iterations. %
Note that we did not complete training on our entire training set; 500 training iterations was sufficiently long to collect profiling data while training was stable.
Finally, since ResNet-50 is commonly trained using multiple GPUs,
we trained across four GPUs on one node and trained 3$-$4 times per node.
Like SGEMM, before collecting data we perform one warm up run.

Unlike SGEMM, ResNet-50 has $\approx$85 unique kernels and over 1.3 million kernels per run.
Since 75\% of these kernels run for less than 2ms, accurately measuring them is challenging (Section~\ref{sec:method}).
Thus, averaging kernel duration would not fairly characterize its overall performance.
Hence, for ResNet-50 we use iteration duration instead of kernel duration as our performance metric; iteration duration is also more informative for HPC users training ML models.
All other metrics are the same as SGEMM.
However, we ignore the initialization kernels (e.g., NCCL) to avoid one-time startup costs.

Figure~\ref{fig:resnet-tacc-summary} presents ResNet-50's aggregated box plots on Longhorn for our 4 metrics.
Unlike our SGEMM experiments on Longhorn (Section~\ref{sec:res-longhorn}), ResNet-50 has little frequency variation and most nodes run at the max 1530MHz.
However, ResNet-50 has 22\% performance variation (Figure~\ref{fig:resnet-tacc-summary}b), our largest observed variation.
The temperature variation (Figure~\ref{fig:resnet-tacc-summary}d) is similar to %
SGEMM's: a 30\degree C range.
Surprisingly, unlike SGEMM there is significant power variation (104\%, Figure~\ref{fig:resnet-tacc-summary}c).
We believe this %
reflects ResNet-50's more varied behavior across its different kernel types -- although its convolution kernels are similar to SGEMM, ResNet-50's other, less computationally intense kernels have less onerous power demands.
Consequently, the DVFS algorithm does not need to reduce voltage or frequency %
to remain beneath the TDP.

Figure~\ref{fig:resnet-tacc-correl} presents scatter plots of our distinct metrics.
Figure~\ref{fig:resnet-tacc-correl}a shows that most iterations complete within 100-150ms and GPUs have frequencies around 1530MHz.
However, surprisingly, there are tails on both axes.
In particular, several c008 runs have much lower frequency yet perform similarly to those running at the max 1530MHz.
Meanwhile, several GPUs in c002 run at 1530MHz (Figure~\ref{fig:resnet-tacc-correl}a), but perform poorly and consume much less power (Figure~\ref{fig:resnet-tacc-correl}b).
We expect c002's stragglers to consume more power than the tail of c008's runs, since $P_{dynamic}$ $\propto frequency$.
However, we observe the opposite: c002's worst performing runs consume much less power (as low as 76W) than c008 and other cabinets.
This differs from Longhorn's SGEMM runs.
Thus, ResNet-50 and SGEMM's variability and correlation differences on Longhorn suggest that variability is application-specific.
Moreover, 8 of the 10 worst-performing GPUs for SGEMM were also outliers for ResNet, highlighting that variability is not transient and some GPUs consistently do not perform well.

\begin{figure*}[tb!]
    \centering
    \includegraphics[width=\textwidth]{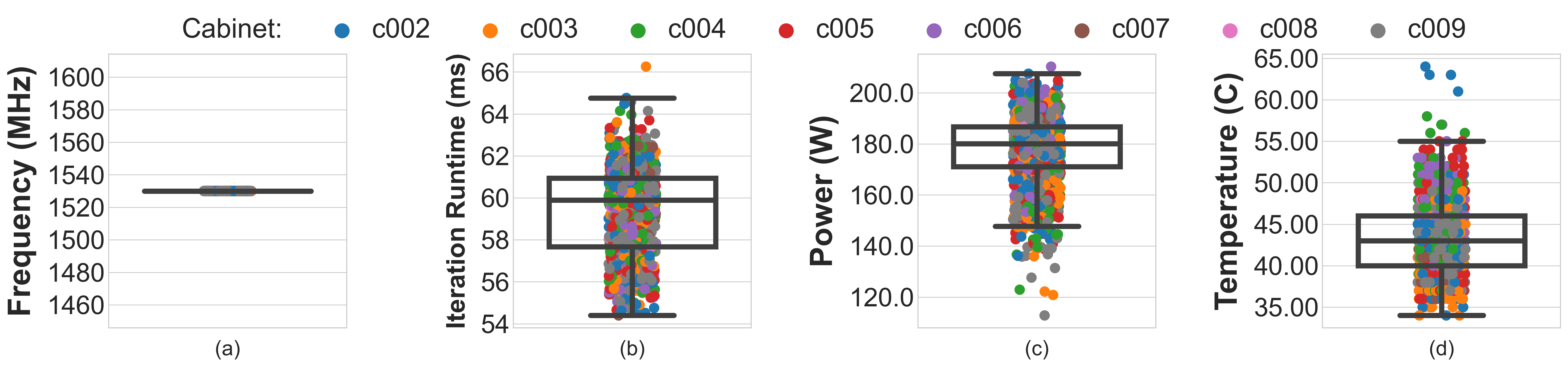}
    \vspace{-0.6cm}
    \caption{Single-GPU ResNet-50 box plot summaries for (a) frequency, (b) performance, (c) power, and (d) temperature.}
    \label{fig:resnet-single-gpu-summary}
    \vspace{-2ex}
\end{figure*}

We also ran ResNet as a single-GPU experiment with the same training set but proportionally scale the batch size to 16.
Figure~\ref{fig:resnet-single-gpu-summary} presents the aggregated boxplots for the single-GPU experiments.
Similar to the multi-GPU runs, power consumption stays well within TDP limits for all GPUs, hence they run at the max frequency of 1530MHz and observe little PM interference.
The absolute power consumption values are lower than multi-GPU experiments, but we still see a significant median power variation (24\%) across GPUs.
Similarly, the absolute iteration duration values are also lower than the multi-GPU results, but single-GPU ResNet continues to demonstrate significant performance variability (14\%).
While the c002 outliers observed for SGEMM and multi-GPU ResNet still run at high temperatures (Figure~\ref{fig:resnet-single-gpu-summary}(d)), they demonstrate lower degradation in performance compared to multi-GPU ResNet.
This shows that multi-GPU jobs with a bulk synchronous pattern end up running as fast as the slowest GPU, leading to larger performance degradation.

\noindent\textbf{Takeaway 5}: \emph{ResNet-50 exhibits the highest performance variability (22\% for multi-GPU, 14\% in single-GPU) across all our benchmarks. Moreover, the difference in ResNet-50's and SGEMM's compute intensities, and the significant difference in performance-frequency correlations, suggests that variability is application-specific.}

\vspace{-1ex}
\subsection{BERT on TACC Longhorn}
\label{sec:res-bert-longhorn}

\begin{figure*}[tb!]
    \centering
    \includegraphics[width=\textwidth]{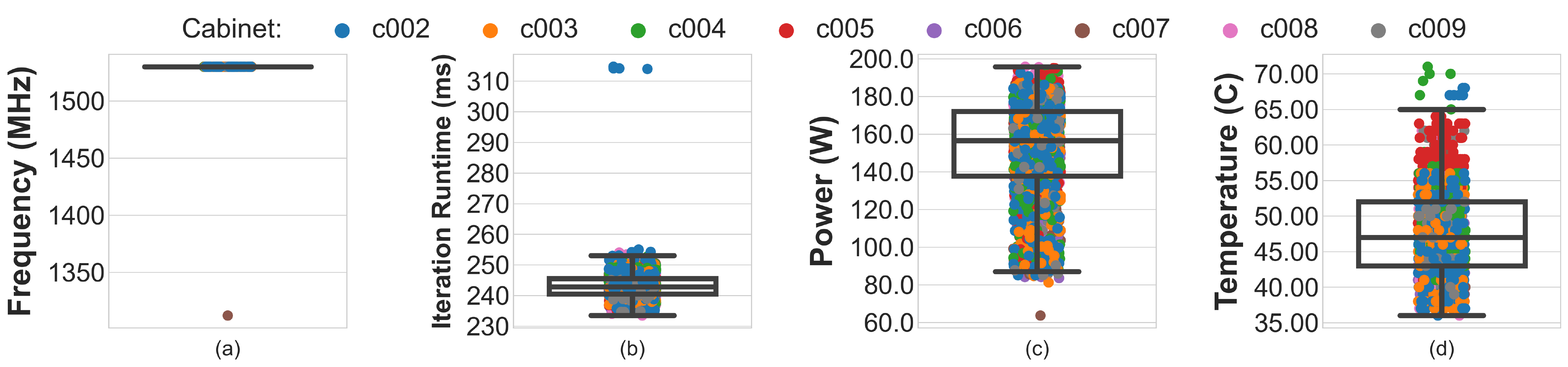}
    \vspace{-0.6cm}
    \caption{Multi-GPU BERT box plot summaries on Longhorn for (a) frequency, (b) performance, (c) power, and (d) temperature.}
    \label{fig:bert-tacc-summary}
    \vspace{-3ex}
\end{figure*} 

Given the increasing adoption of Transformer-based models~\cite{BahdanauCho2015-nmt, correia2019adaptively, JouppiYoon2021-tpuv4, kitaev2020reformer, LanChen2019-albert, LuongPham2015-attentionNMT, ShenZhou2018-bidirAttention, ShoeybiPatwary2019-megatronlm, WuSchuster16-seq2seq, ZhuXia2020-bertNMT}, we also study another multi-GPU ML workload: pre-training of Bidirectional Encoder Representations from Transformers (BERT)~\cite{DevlinChang18-bert}.
BERT is widely used in Natural Language Processing (NLP) %
and is in the popular MLPerf suite~\cite{MattsonCheng2020-mlPerfTraining, MattsonReddi2020-mlPerf}.
We used BERT$_{LARGE}$, which has 24 encoders with 16 bidirectional self-attention heads.
Our training set was 30522 words and uses a batch size of 64.
Like ResNet, we limit each training run to 250 iterations and run across all 4 GPUs in a node.
We performed one warm up run and then performed measurements 5 times on each node. 
Overall we observed 53 nodes in Longhorn for BERT and collect the same metrics as ResNet.
Similar to ResNet, we omit the initialization kernels to avoid one-time startup costs.

Figure~\ref{fig:bert-tacc-summary} shows aggregated box plots on Longhorn for our 4 metrics.
Compared to ResNet, BERT has less overall power consumption (Figure~\ref{fig:bert-tacc-summary}c): BERT's median power consumption is around 40W lower.
This is expected because ResNet uses more compute intensive GEMMs, which increase its power consumption.
In comparison, BERT's GEMMs are much less computationally intensive: although GEMMs make up 30-65\% of its total runtime, they only utilize 40-50\% of the GPU~\cite{IvanovDryden2021-transformersDataMove, PatiAga2021-demystifying, ShoeybiPatwary2019-megatronlm, ZadehPoulos2019-dlTime} -- which decreases BERT's overall power consumption.
Nonetheless, like ResNet there is large power variability ($\approx$87$\%$).
However, because BERT is less computationally intensive than ResNet, it has less performance variability ($\approx$8$\%$) and fewer frequency outliers (Figure~\ref{fig:bert-tacc-summary}b and Figure~\ref{fig:bert-tacc-summary}a, respectively).
Perhaps unsurprisingly, the performance outliers in Figure~\ref{fig:bert-tacc-summary} in cabinet c002 on Longhorn are also outlier nodes for ResNet in Figure~\ref{fig:resnet-tacc-summary}b.
This suggests that the same GPUs perform poorly for both ML applications.

\noindent\textbf{Takeaway 6}: \emph{Like ResNet-50, BERT exhibits large power variability.  However, BERT's less computationally intensive GEMMs reduce performance variability (8\%).  Moreover, BERT's and ResNet-50's outlier nodes are the same.}

\vspace{-1ex}
\subsection{LAMMPS on TACC Longhorn}
\label{sec:res-lammps-longhorn}

\begin{figure*}[pbth!]
    \centering
    \includegraphics[width=\textwidth]{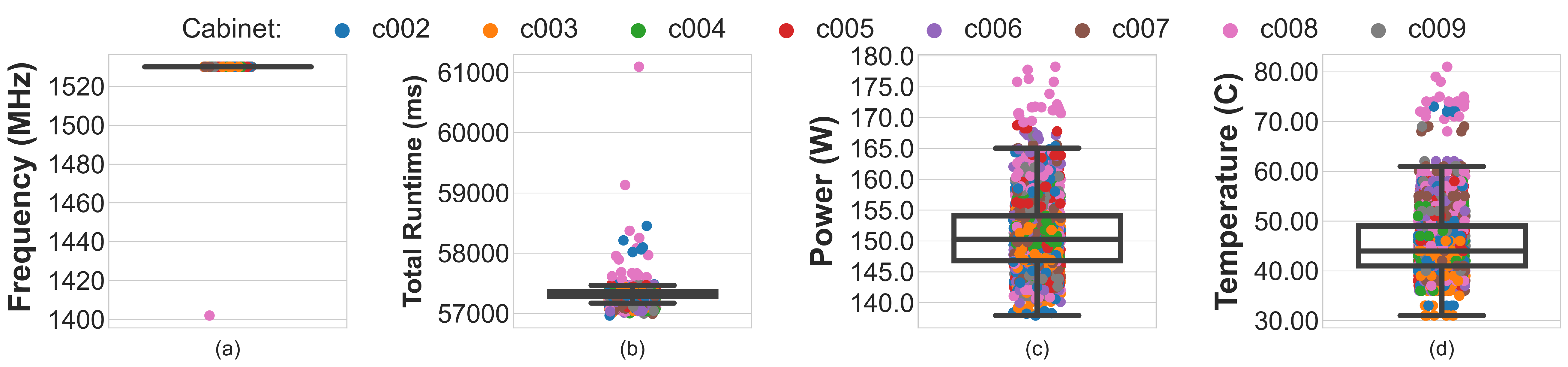}
    \vspace{-0.6cm}
    \caption{LAMMPS box plot summary results on Longhorn for (a) frequency, (b) performance, (c) power, and (d) temperature.}
    \label{fig:lammps-tacc-summary}
    \vspace{-2ex}
\end{figure*}

LAMMPS is a popular molecular dynamics (MD) application that simulates a variety of atomic and molecular systems~\cite{LAMMPS}.
Whether LAMMPS is compute or memory-bound depends on the selected settings and hardware~\cite{coral2,Vyacheslav-lammpsOnJetson,Aktulga2012-ReaxFF}.
We chose to use the REAXC setting~\cite{Shan2015-ReaxFFSource} to simulate a chemical reaction. 
When REAXC is run in a distributed multi-GPU setup with large simulation sizes, LAMMPS is
compute-intensive~\cite{coral2}.
However, to compare LAMMPS variability with SGEMM, we ran it as a single-GPU experiment, with input configuration parameters $(x,y,z)$ that determine GPU occupancy and are limited by the 16GB device memory on NVIDIA V100s. After careful tuning, we selected the (8,16,16) input size, which lead to high GPU utilization, while still remaining within device memory limits.
Profiling shows that LAMMPS has 42$\times$ higher DRAM utilization than ResNet, while ResNet kernels utilized FUs 4.3$\times$ more than LAMMPS.
Thus, LAMMPS is memory-bound in our experiments.

Each LAMMPS run is composed broadly of 2 types of kernels, short-running ($\leq$60$\mu$s long) and long-running kernels (4 unique kernels, 20-200ms long).
Long-running kernels make up 98$\%$ of the total runtime of a LAMMPS job but there are 4 unique long kernels interspersed with short ones.
Thus, similar to ResNet-50, median kernel duration is not a good performance measure. %
We therefore use the sum of all large kernel durations as our performance metric.
All other metrics are measured as specified in Section~\ref{sec:method}.

\begin{figure*}[tb!]
    \centering
    \includegraphics[width=\textwidth]{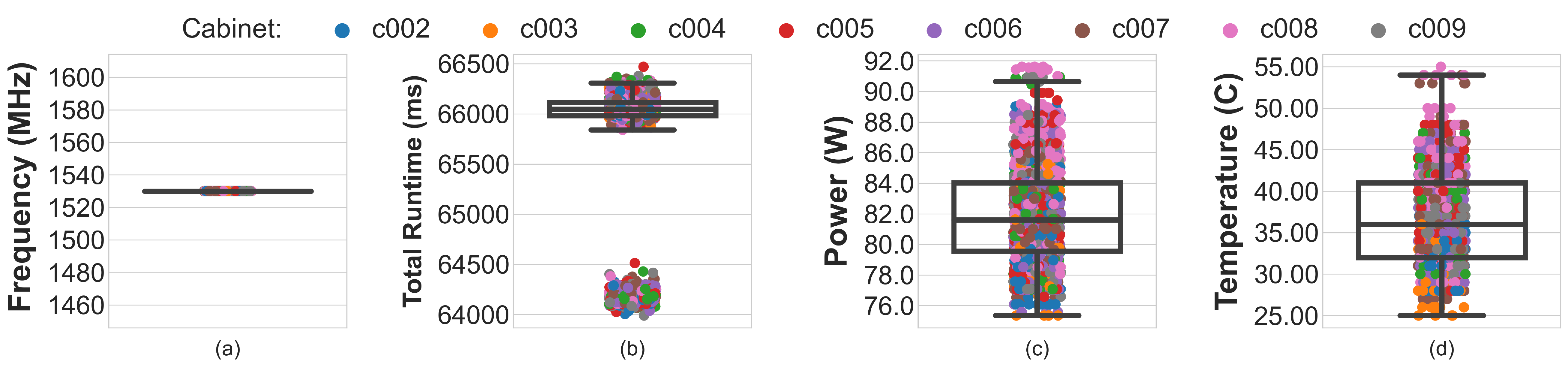}
    \vspace{-0.6cm}
    \caption{PageRank box plot summary results on Longhorn for (a) frequency, (b) performance, (c) power, and (d) temperature.}
    \label{fig:pagerank-tacc-summary}
    \vspace{-3ex}
\end{figure*}

Figure~\ref{fig:lammps-tacc-summary} presents LAMMPS' aggregated box plots on Longhorn for our 4 metrics.
Interestingly, median power for all LAMMPS jobs was $\leq$180 W (Figure~\ref{fig:lammps-tacc-summary}c), in contrast to more compute-intensive applications such as SGEMM, which often touched V100 TDP.
Moreover, similar to BERT and ResNet-50, frequency quickly saturates to the maximum value of 1530MHz (Figure~\ref{fig:lammps-tacc-summary}a) and does not change throughout the course of the job.
Additionally, performance varies by less than 1\% in Figure~\ref{fig:lammps-tacc-summary}b, which completely differs from BERT (8\%), SGEMM (9\%), and ResNet-50 (22\%).
However, we still observe power variability of 20\% and temperature variability
of 8\degree C between $Q1$ and $Q3$ (Figure~\ref{fig:lammps-tacc-summary}d).
These observations are in line with prior work~\cite{Vyacheslav-lammpsOnJetson} and emphasize that high energy consumption is undesirable in a memory-bound application because it is not accompanied by any significant increase in performance.
In part, this happens because the GPU's memory frequency is lower than the compute frequency, and even if the application is memory-bound, this does not stress the TDP as much as the compute-heavy applications.
Overall, this suggests that (i) SM frequency gets pinned in applications that
are less compute-intensive (ii) performance of such applications is relatively more predictable with very low variability in application runtime and (iii) significant temperature and power variability are still observed across GPUs.

\noindent\textbf{Takeaway 7}: \emph{Similar to \textbf{Takeaway 5}, LAMMPS' results show that variability is application-specific with memory-intensive applications seeing lower performance variance, but still having significant power and temperature variability.}

\vspace{-1ex}
\subsection{PageRank on TACC Longhorn}
\label{sec:res-pagerank-longhorn}
\vspace{-1ex}

\begin{figure*}[tb!]
  \centering
  \begin{subfigure}{.45\textwidth}
    \centering
    \includegraphics[width=\textwidth]{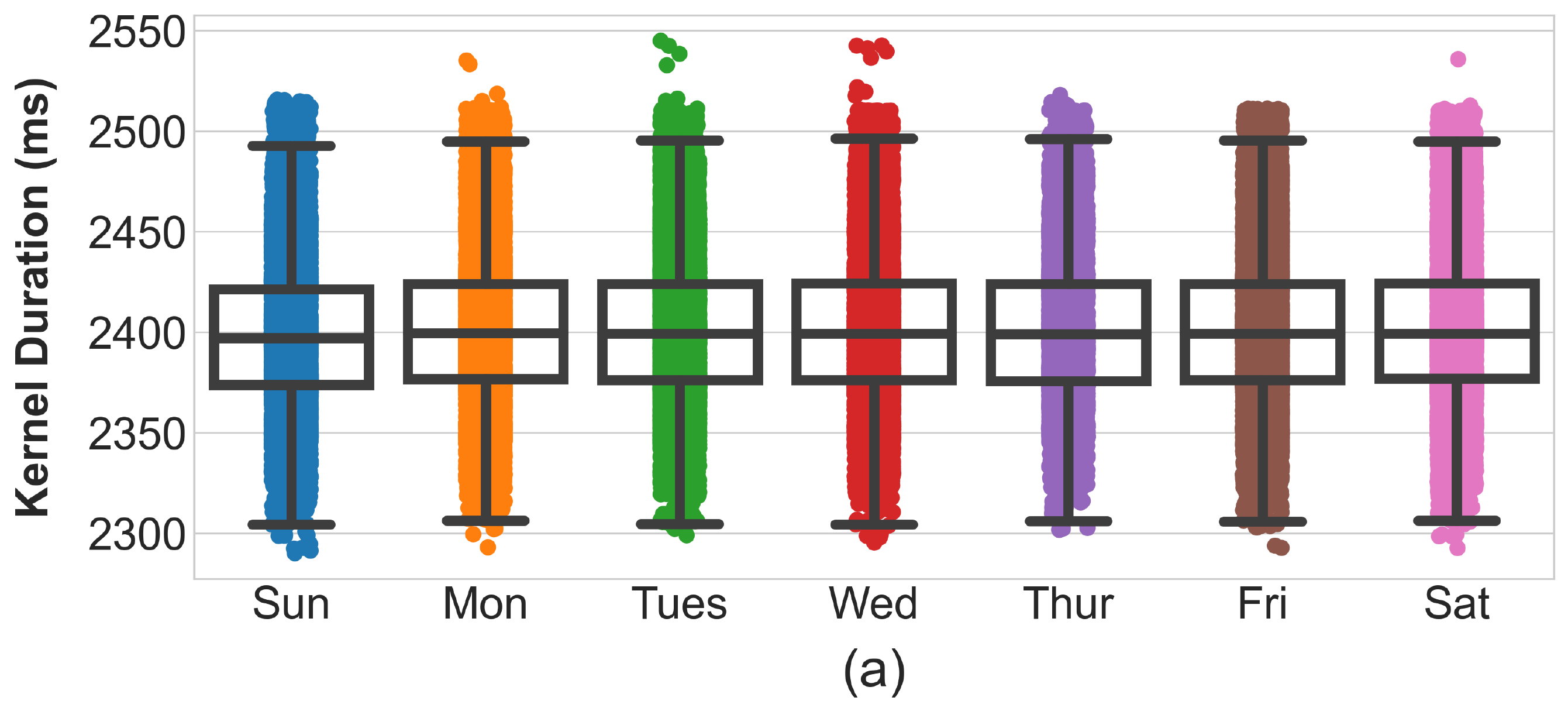}
    \label{fig:summit-weekday-perf}
  \end{subfigure}
  \hspace{0.3in}
  \begin{subfigure}{.45\textwidth}
    \centering
    \includegraphics[width=\textwidth]{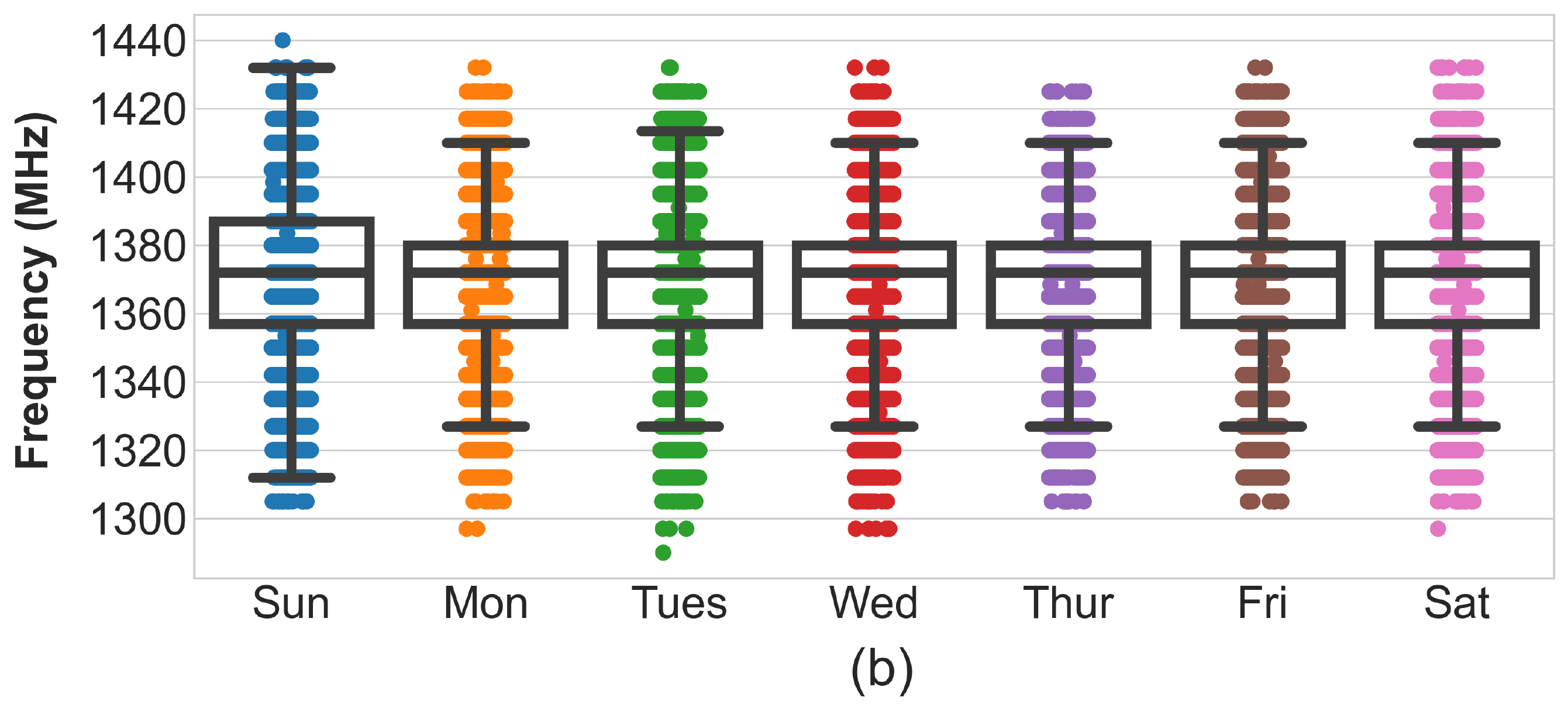}
    \label{fig:summit-weekday-freq}
  \end{subfigure}
  \vspace{-0.1in}
  \begin{subfigure}{.45\textwidth}
    \centering
    \includegraphics[width=\textwidth]{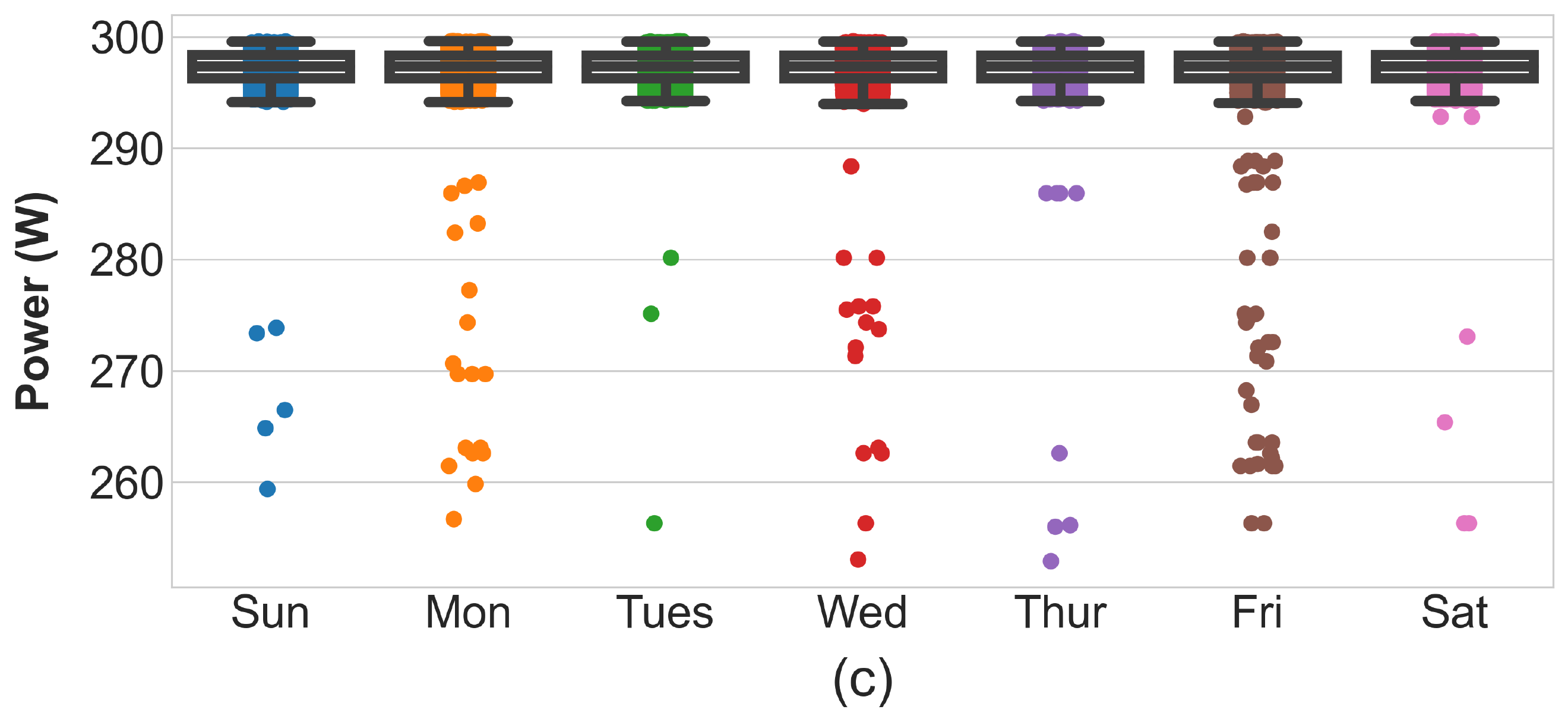}
    \label{fig:summit-weekday-pwr}
  \end{subfigure}
  \hspace{0.3in}
  \begin{subfigure}{.45\textwidth}
    \centering
    \includegraphics[width=\textwidth]{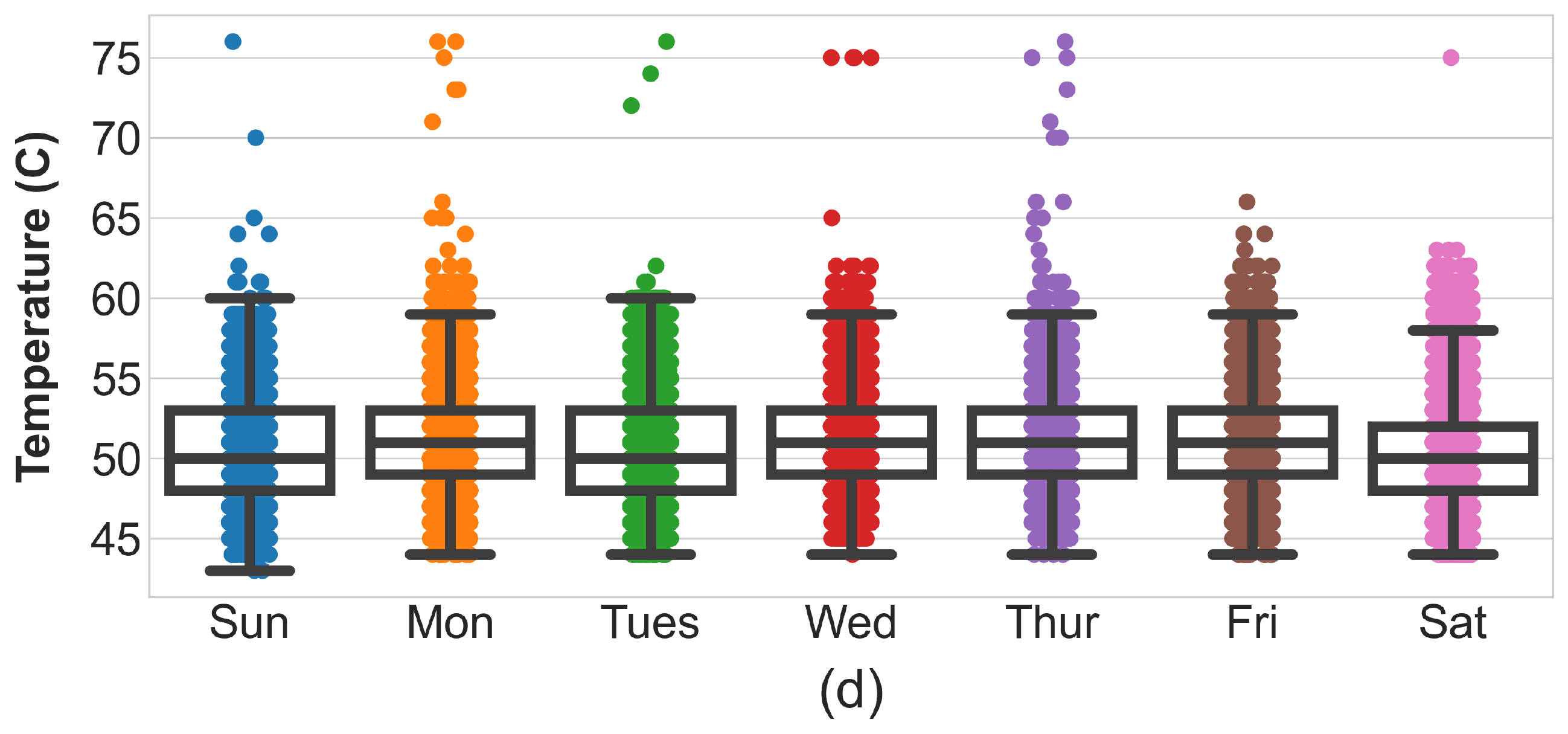}
    \label{fig:summit-weekday-temp}
  \end{subfigure}
  \vspace{-0.3cm}
  \caption{Day of the week summary results for Summit. We see more power outliers on Mondays, Wednesdays, and Fridays.}
  \label{fig:summit-weekday-summary}
  \vspace{-1ex}
\end{figure*}

PageRank is a popular graph analytics algorithm that is used in recommender systems, search engines, social network analysis, and bibliometrics~\cite{PageRankSpMV}.
Although neither push- nor pull-based graph analytics algorithms always provide the best performance on GPUs~\cite{BestaPodstawski2017-pushPullGraphAnaly, SalvadorDarvin2020-gpuGraphs, SorensenPai2019-gpuGraphs}, we focus on pull-based algorithms because they are more widely used.
The most common PageRank algorithm utilizes Sparse matrix-Vector (SPMV) computation~\cite{PageRankSpMV, CheBeckmann2013-pannotia}.
Since PageRank's sparsity is input graph dependent, its memory access pattern can be highly irregular and often dwarfs the amount of compute, making it both memory bandwidth-bound and highly irregular. 
We chose an input graph that fully utilizes the SMs of a V100 GPU and provides sufficiently long kernels runtimes %
(Section~\ref{sec:method}).
We ran PageRank with {\it rajat30}, an undirected graph for circuit simulation~\cite{rajat30}.
The other configuration parameters are the same as SGEMM on Longhorn.
Like LAMMPS, SPMV computations are also memory bound, but irregular.  Consequently, they do not stress memory as much: LAMMPS has 4.24$\times$ higher DRAM utilization than PageRank.  However, PageRank kernels had $61\%$ memory dependency stalls, versus $7\%$ for LAMMPS and $3\%$ for SGEMM. PageRank also had negligible FU execution dependency stalls ($12\times$ less than SGEMM), showing that PageRank is not compute-bound.

Figure~\ref{fig:pagerank-tacc-summary} summarizes Longhorn variability for PageRank.
Similar to LAMMPS (Section~\ref{sec:res-lammps-longhorn}), PageRank has little frequency variation and very little performance variability (1\%) across GPUs.
However, we observed 22$\%$ variability in median power across runs and a temperature variation of 8\degree C between $Q1$ and $Q3$, as seen with LAMMPS.

\noindent\textbf{Takeaway 8}: \emph{Like LAMMPS (\textbf{Takeaway 6}), PageRank has little frequency and performance variation. This suggests that memory-bound workloads can use worse-performing nodes without significant performance penalty.}

\section{Variation across time, power budget}
\label{sec:res-variation-time}

Next, we analyze how variability changes across days of the week and how the GPU power limit affects variability.

\begin{figure*}[tb!]
  \centering
    \includegraphics[width=\textwidth]{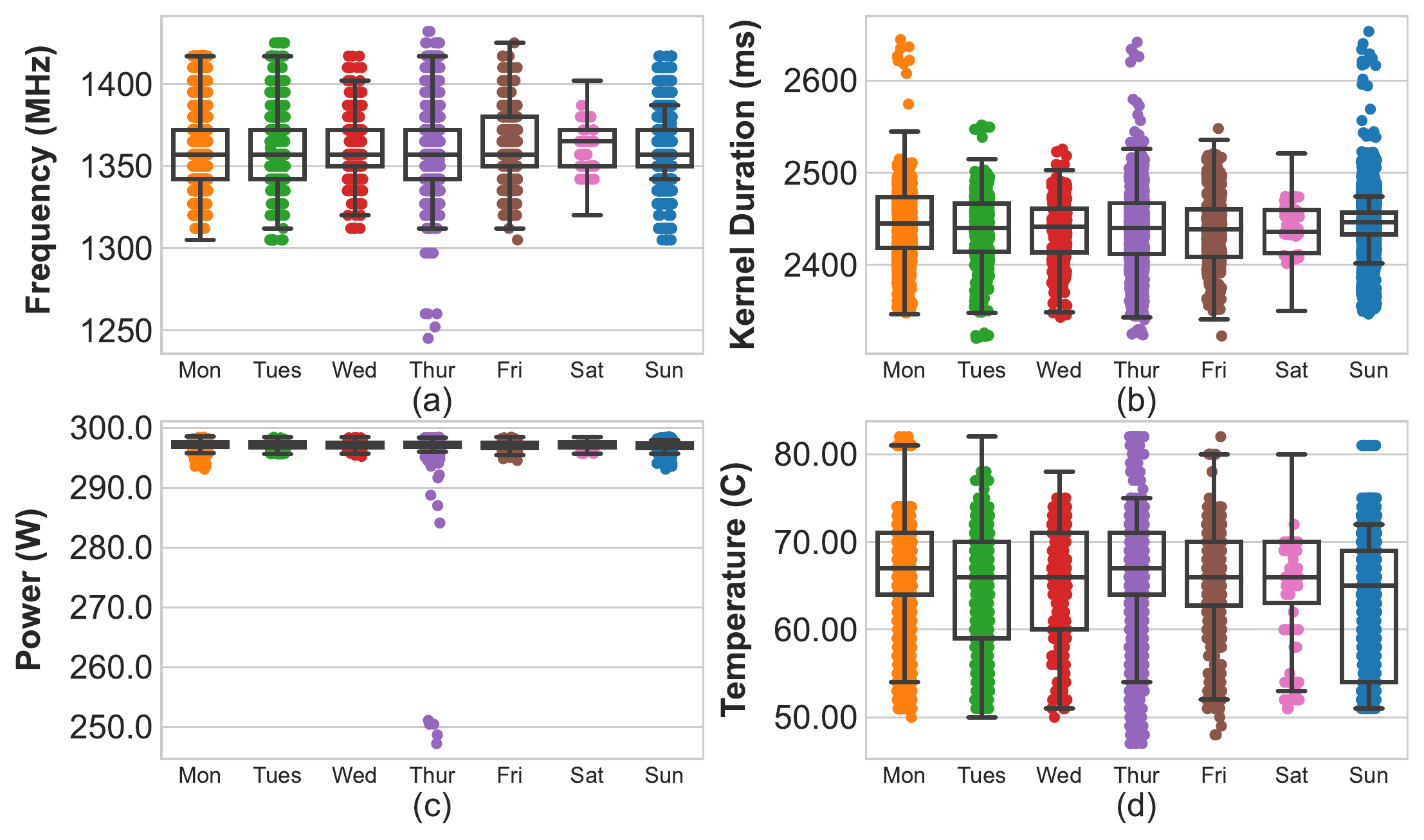}
  \vspace{-0.4cm}
  \caption{Day of the week summary results for Longhorn. We observed performance outliers on Mondays, Thursdays and Sundays.}
  \label{fig:longhorn-weekday-summary}
  \vspace{-4ex}
\end{figure*}

\subsection{Variation Across Days of the Week}
\label{subsec:res-dayOfWeek}

To determine if the variations we observed on these clusters hold over time, we ran SGEMM on Summit on each day of the week across a period of eight weeks.
Figure~\ref{fig:summit-weekday-summary} shows a consistent trend across each day of the week: around 8\% variation.
On Mondays, Wednesdays, and Fridays there is a higher concentration of GPUs which consume below 290W.
However, even with more power outliers, performance on these days is very similar to the rest of the
week and \textbf{Takeaways 1-3}, as is also shown in Figure~\ref{fig:summit-scatterplots}.

Longhorn also shows similar trends, as shown in Figure~\ref{fig:longhorn-weekday-summary}. There is consistent performance variability across all days of the week. Compared to Summit, however, the variation is lower - around 3\% each day. We also observe more outliers on Mondays, Thursdays and Fridays than other days of the week. 

\noindent\textbf{Takeaway 9}: \textit{The variability we observe is consistent throughout the week, suggesting that regardless of when the experiments are run, our observations hold.}

\subsection{Varying Power Limit}
\label{subsec:powerlimit}

\begin{figure}[tb!]
    \centering
    \includegraphics[width=\columnwidth]{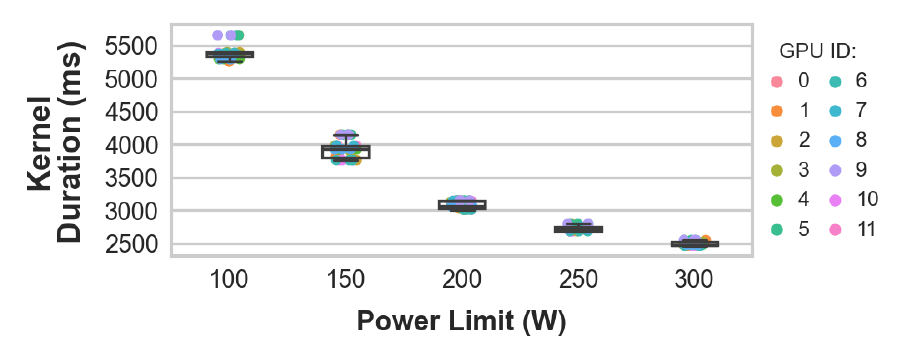}
    \vspace{-0.6cm}
    \caption{Performance variation of SGEMM on NSF CloudLab GPUs when varying the power limit.}
    \label{fig:power-limit-summary}
    \vspace{-4ex}
\end{figure}

We also studied how varying a GPU's power limit affects variability.
Varying the power limit (e.g., using \texttt{nvidia-smi}) requires administrative access.
Since we did not have administrator (root) access on the large compute clusters, we used the smaller CloudLab~\cite{DuplyakinRicci2019-cloudlab}.
As discussed in Section~\ref{sec:method}, CloudLab has 12 NVIDIA V100 GPUs.
Figure~\ref{fig:power-limit-summary} shows the performance variation as the power limit varies from 100-300W while running SGEMM.
Interestingly, unlike prior work~\cite{Scogland2015-pwrPerspectives} performance variation was similar when pinning and not pinning (Section~\ref{sec:res-longhorn}).
Rather, our variability results for a 300W limit are in line with those from the large clusters: 9\% variation.
Moreover, as expected, the kernel durations increase with lower power limits.
However, variability and the number of outliers also increase with lower power limits.
For example, at a 150W limit there is 18\% variability versus 9\% at a 300W limit.
Our conversations with GPU manufacturers indicate that this is potentially occurring because GPU DVFS algorithms are less optimized for extremely low power budgets, but variability under power limits would become important if future exascale machines are operating under a varying power budget~\cite{power-limit}.

\section{Conclusions and Takeaways}
\label{sec:takeaways}

Modern compute clusters are embracing accelerator-rich systems, especially GPUs.
Although prior work has identified how variability in CPUs affects these systems, it is unclear how much GPU variability affects these systems.
Thus, we conducted a detailed study and analysis of how GPU variability affects a wide range of modern HPC and scientific workloads across five computing centers of varying scales, cooling methods, and GPU vendors.
Our results show that there is significant variability in these systems: up to 22\% for the applications, with outliers up to \textbf{1.5$\times$} slower than the median GPU.
This demonstrates the need to embrace variability in future accelerator-based systems.
We conclude by highlighting several implications and mitigation strategies that practitioners and researchers can pursue in the future: %

\noindent \textbf{Impact on Users}:
In addition to the performance variation, inter-experiment variability is also important.
For example, when running SGEMM on Longhorn, %
18\% of the GPUs 
are 6-7\% (or about 150ms) slower than the fastest GPUs.
Thus, users running SGEMM-like single GPU experiments would have an 18\% chance of being assigned a slower GPU.
In Summit, again with SGEMM, %
9\% of the GPUs are 6-7\% (or about 160ms) slower than the fastest GPUs. %
Thus, users running a single GPU experiment on Summit have a 9\% chance of being assigned a slower GPU.
Like Longhorn, Summit also has variance within specific GPUs.
Users running multi-GPU experiments are even more likely to be assigned a slower GPU: if a user requests 4 GPUs on Longhorn, 40\%-50\% of the time they will be assigned a slower GPU.%
We also observed this in our multi-GPU BERT and ResNet-50 experiments.

\noindent
\textbf{Spatial Effects}: For all of our experiments, we obtained exclusive access to a machine and all its GPUs.
This eliminated any spatial effects from jobs running on neighboring GPUs.
While this is the typical allocation mode in modern supercomputing clusters, spatial effects would be relevant for other scenarios like cloud computing or enterprise clusters~\cite{jeon2019analysis} where GPUs are allocated individually.
We plan to study both spatial and temporal (i.e., variability due to a preceding job run on the same GPU) effects in the future.

\noindent
\textbf{Blacklisting, Maintenance}:
Cluster operators can use our study to improve the cluster's operation and help develop strategies for better maintenance.
For example, our study has already helped TACC's operators identify and perform targeted maintenance on problematic nodes with clearly underperforming GPUs in Frontera and Longhorn.
Performing periodic variability benchmarking can help automate this.

\noindent
\textbf{Application-aware Frameworks}: 
Since performance variation is application dependent, the next generation of HPC allocation frameworks should take application properties into account to mitigate variance.
Similar to prior work~\cite{Guerreiro-appClasses}, our profiling indicates that metrics like FU utilization, DRAM utilization, and memory stalls can be used by operators to classify applications and modify schedulers to assign medium- and high-compute intensity workloads on nodes with less variation.
Similarly, memory-bound applications can be run on higher-variation nodes without incurring significant performance loss.
However, this may change in future as thermal performance degrades below 14nm~\cite{DRAMthermalissues}.

\noindent
\textbf{New Hardware and System Design}:
A major limiter to further improving efficiency is the lack of standards for exposing power information in modern accelerators.
Thus, for future systems, designers can build on the insights generated by our benchmarks for current systems and apply co-design that makes the hardware, software, and runtime layers aware of the variance in the systems.
To do this, we will need to design a standard for accelerators to expose PM information from the hardware to the software and runtime. Using this information, we can develop techniques for \textbf{global power management} that can enable optimal PM decisions across accelerators and further reduce performance variability.

\section*{Acknowledgments}

We thank the anonymous shepherd and the SC reviewers for their constructive comments and suggestions that improved this work.
We also thank Zhao Zhang (TACC), Clayton Hughes (SNL), Oscar Hernandez and David Bernholdt (ORNL), Ian Karlin and Scott Futral (LLNL) for providing us machine access on the clusters we used and helping us run some of the experiments.
Additional thanks Ian and Scott for providing information about slowdown temperatures for the Corona GPUs.
We extend our gratitude to Stan Moore and Christian Trott (SNL) for their help with using Kokkos library for LAMMPS.
We also acknowledge Scott Groel at Cloudlab Clemson for clarification on V100 fans.
Finally, we thank Joseph Greathouse (AMD) and Mike Sabotta (NVIDIA) for helping us understand GPU power management.
This work is supported in part by a University of Wisconsin Fall Research Competition grant, a LLNL allocation, ORNL grant GEN010wisc, and a TACC allocation.
Sinclair has an affiliate appointment with AMD Research.

\bibliographystyle{IEEEtran}
\bibliography{refs}

\newpage
\newpage
\begin{appendices}
\label{sec:apdx}

\section{Artifact Description/Evaluation}
\label{sec:apdx-artifact}

\subsection{Abstract}
\label{subsec:apdx-artifact-abs}

This artifact provides container specifications and scripts for reproducing experiments that measure and benchmark variability of GPUs across different applications. The artifact can be used to run different benchmarks across machine learning, molecular dynamics and graph analytics. We also include scripts that use vendor-specific profiling tools to measure power (W), temperature ($\degree$ C), frequency (MHz) and performance (kernel runtime/iteration duration/total runtime depending on application). The experiments, their corresponding paper sections and the base framework/tools used to run them are summarised in Table~\ref{tab:summary-adae}.

\begin{table}[h!]
    {\footnotesize
  \centering
  \begin{tabular}{|c|c|c|}
    \hline
    \multirow{2}{*}{\textbf{Experiment}} & \multirow{2}{*}{\textbf{Section}} & \textbf{Base Framework/} \\
    & & \textbf{Tools Used} \\ \hline
    \textbf{SGEMM on} & \multirow{2}{*}{
    \ref{sec:res-sgemm-methodology}} & CUDA 10.1, nvcc 10.1 \\ 
   \textbf{NVIDIA GPUs}\cite{cublas} & & NVProf 10.1, gcc 4.8.5 \\ \hline
    \textbf{SGEMM on} & \multirow{2}{*}{
    \ref{sec:res-sgemm-corona}} & \multirow{2}{*}{ROCm 4.0.1} \\ 
   \textbf{AMD GPUs}\cite{cublas} & & \\ \hline     
    \multirow{2}{*}{\textbf{ResNet-50}\cite{ResNet-pyTorchRef}} & \multirow{2}{*}{ 
    \ref{sec:res-resnet-longhorn}} & CUDA 10.1, NVProf 10.1 \\
     & &  PyTorch 1.9  \\ \hline
    \multirow{2}{*}{\textbf{BERT}\cite{DevlinChang18-bert}} & \multirow{2}{*}{ \ref{sec:res-bert-longhorn}} & CUDA 10.1, NVProf 10.1   \\
    & & PyTorch 1.9 \\ \hline
     \multirow{2}{*}{\textbf{LAMMPS}\cite{LAMMPS}} & \multirow{2}{*}{ \ref{sec:res-lammps-longhorn}} & CUDA 10.1, NVProf 10.1,  \\ 
     & & nvcc 10.1 \\ \hline
    \multirow{2}{*}{\textbf{PageRank}\cite{CheBeckmann2013-pannotia}} & \multirow{2}{*}{\ref{sec:res-pagerank-longhorn}}  & CUDA 10.1, NVProf 10.1 \\ 
    & & nvcc 10.1 \\ \hline
  \end{tabular}
  \vspace{-1ex}
  \caption{Summary of experiments in artifact.}
  \label{tab:summary-adae}
  \vspace{-4ex}
}
\end{table}

\subsection{Artifact Metadata}
\label{subsec:apdx-artifact-meta}

\begin{itemize}
    \item \textbf{Persistent ID}: DOI: 10.5281/zenodo.7010207
    \item \textbf{GitHub URL}: \url{https://github.com/hal-uw/gpu_variability_sc22_artifact}
    \item \textbf{Artifact name}: Artifact: Characterizing Variability in Large-Scale, Accelerator-Rich Systems
    \item \textbf{Run-time environment}: Singularity
    \item \textbf{Dependencies}: See Section~\ref{subsubsec:apdx-artifact-prereqs}
\end{itemize}

\subsection{Description}
  The artifact is organized into directories per-experiment (and GPU vendor) to allow any subset of the paper's experiments to be reproduced independent of one another. We include separate singularity scripts for each experiment. These scripts pull container images, install all dependencies and compile library code into respective application binaries. Directions to run each application using these scripts can be found in the respective application's \texttt{README.md} file. At the end of the \texttt{README}s, we also provide troubleshooting/FAQ sections for common errors that users could encounter, and how to resolve them.

\subsubsection{Prerequisites}
\label{subsubsec:apdx-artifact-prereqs}

The following hardware and software dependencies must be satisfied to run the experiments successfully:  
\begin{itemize}
  \item Machine/compute node with an NVIDIA/AMD GPU
  \item Relevant GPU drivers are installed
  \item Singularity is installed 
\end{itemize} 

All the build and run scripts have been tested with Singularity v3.7.2-4.el7a on either NVIDIA V100 (Volta), RTX 5000 (Turing), or AMD MI60 GPUs.

\subsubsection{Installation}
\label{subsubsec:apdx-artifact-install}

The artifact repository can be accessed through the persistent DOI \url{https://doi.org/10.5281/zenodo.7010207}. The main \texttt{README} file in the artifact provides step-wise instructions on cloning the repository and points to per-application instructions for running different experiments. 

\subsubsection{Customization}
\label{subsubsec:apdx-artifact-custom}

Each experiment can be customized for different GPU architectures, different input sizes, and different iterations/time steps. For instance, all of the NVIDIA scripts assume a Volta V100 GPU by default, but the scripts can be configured differently if the user has a different GPU. The \texttt{Prerequisites} section in each application-specific \texttt{README} provides further details on how to customize respective experiments. 

\section{Outlier Analysis: SGEMM on ORNL Summit}
\label{sec:apdx-ornl}

\subsection{Row H Results}
\label{subsec:res-summit-rowh}

\begin{figure*}[p]
  \centering
  \begin{subfigure}{.72\textwidth}
    \centering
    \includegraphics[width=\textwidth]{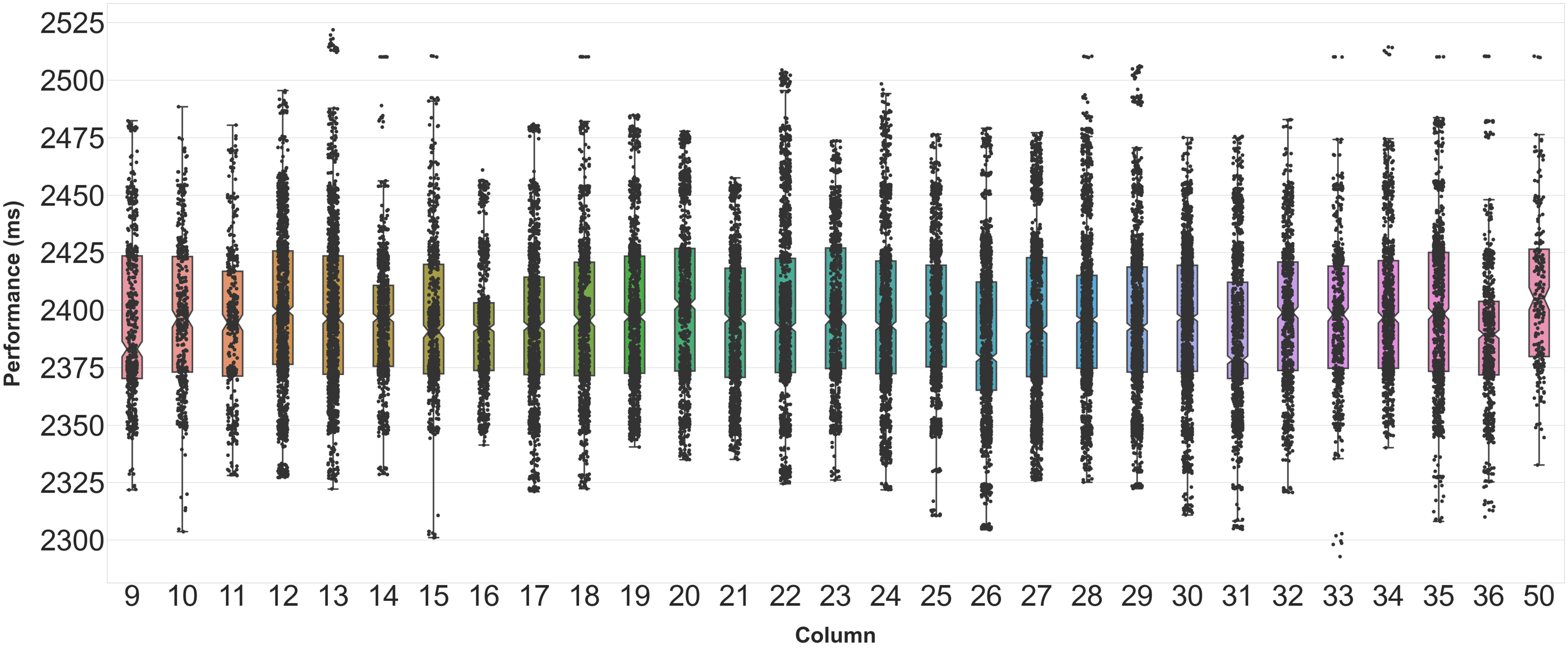}
    \caption{Performance}
    \label{fig:summit-row-h-perf}
  \end{subfigure}
  \begin{subfigure}{.72\textwidth}
    \centering
    \includegraphics[width=\textwidth]{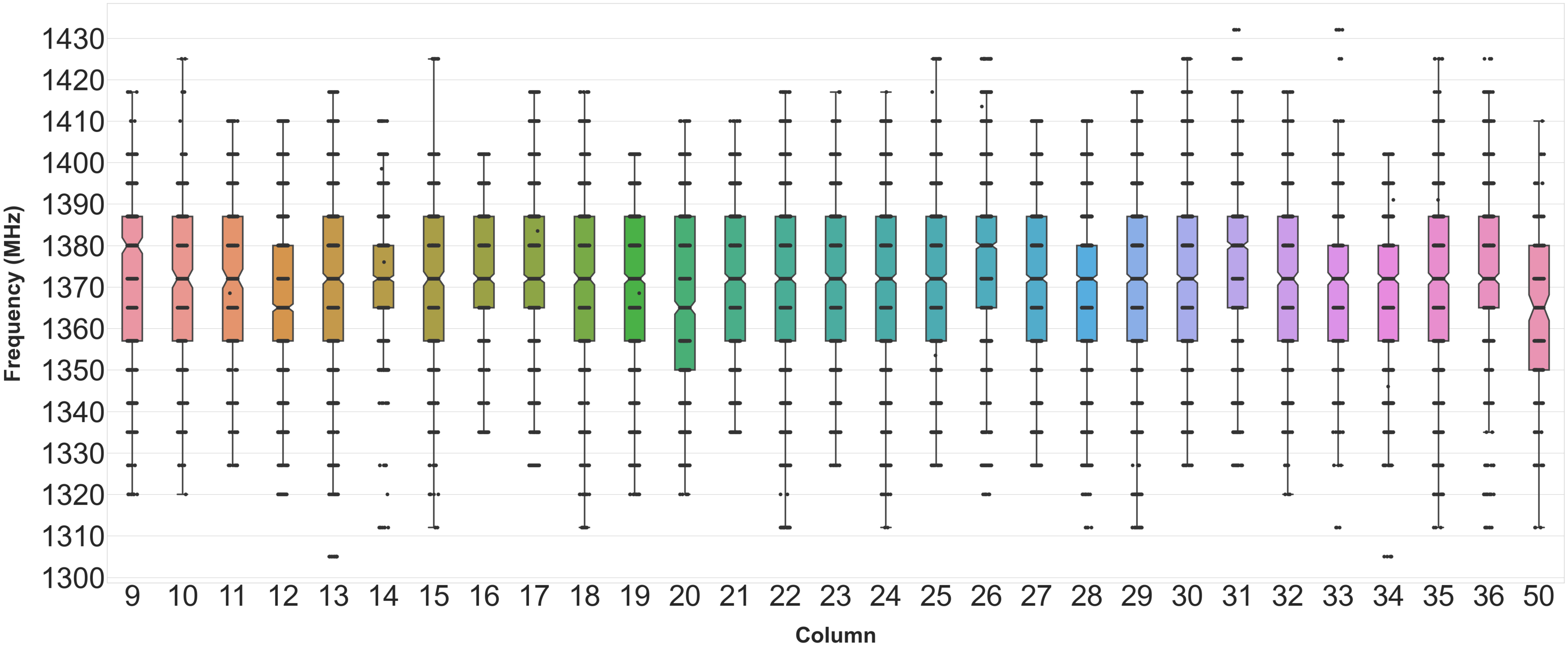}
    \caption{Frequency}
    \label{fig:summit-row-h-freq}
  \end{subfigure}
  \begin{subfigure}{.72\textwidth}
    \centering
    \includegraphics[width=\textwidth]{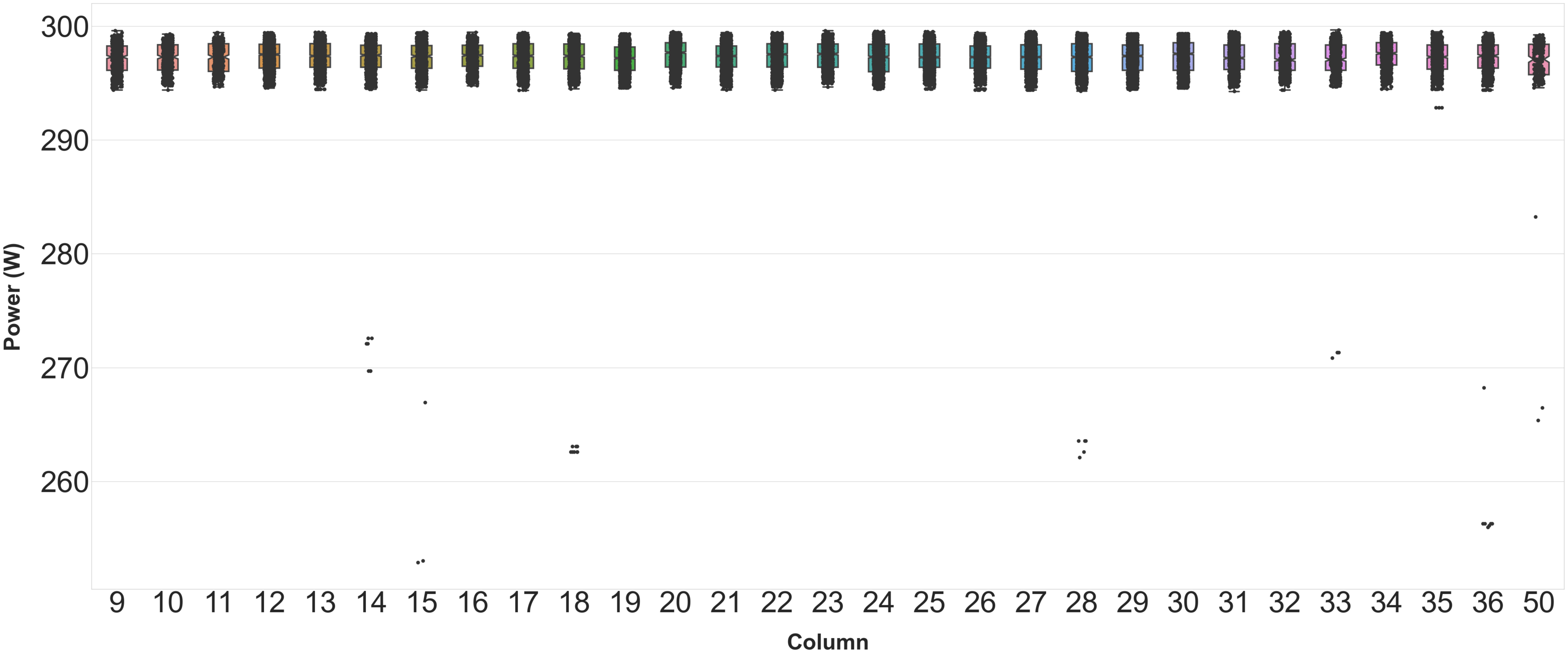}
    \caption{Power}
    \label{fig:summit-row-h-pwr}
  \end{subfigure}
  \begin{subfigure}{.72\textwidth}
    \centering
    \includegraphics[width=\textwidth]{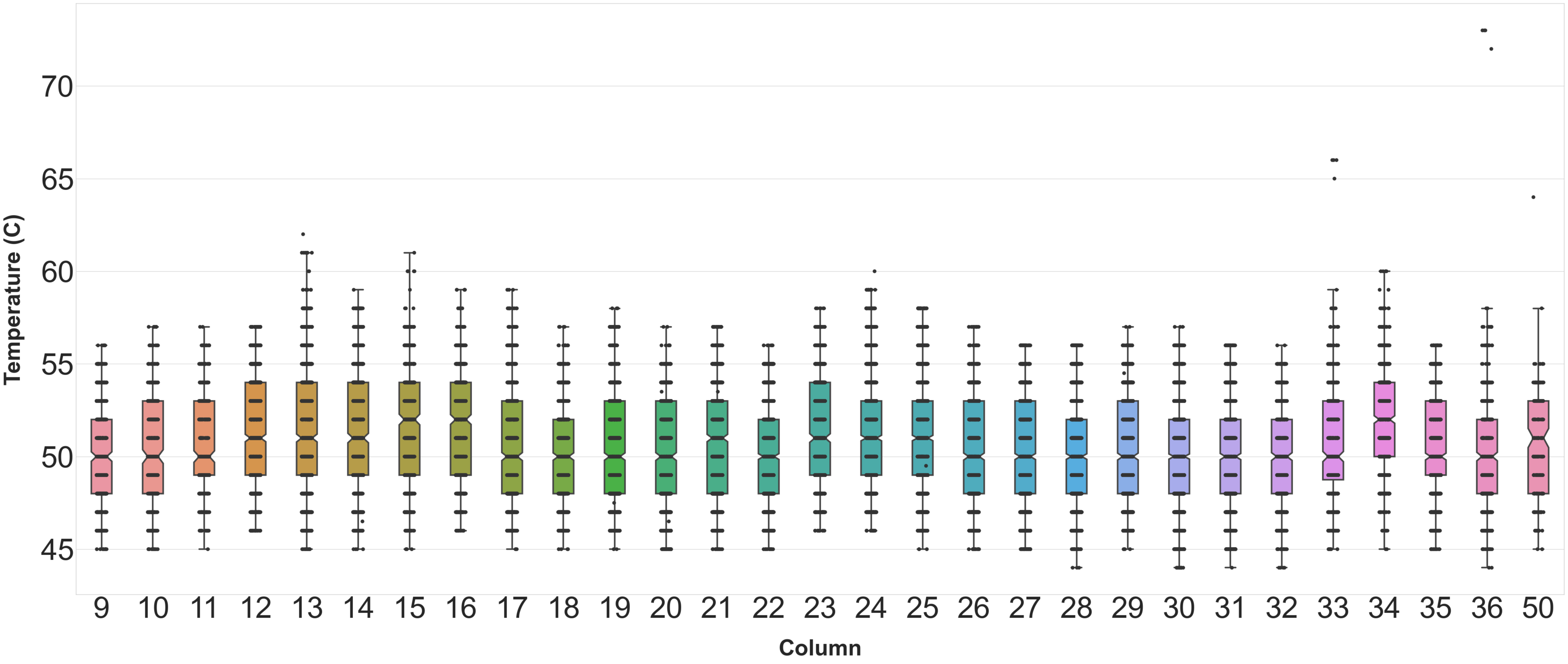}
    \caption{Temperature}
    \label{fig:summit-row-h-temp}
  \end{subfigure}
  \vspace{-0.3cm}
  \caption{Summary results for Summit Row H, showing variation in performance (kernel duration), frequency, power and temperature as reported by NVIDIA's profiler, when running the GPUs unthrottled at the TDP value of 300W.}
  \label{fig:summit-row-h-summary}
\end{figure*}

\begin{figure*}[!t]
  \centering
  \begin{subfigure}{.45\textwidth}
    \centering
    \includegraphics[width=\textwidth]{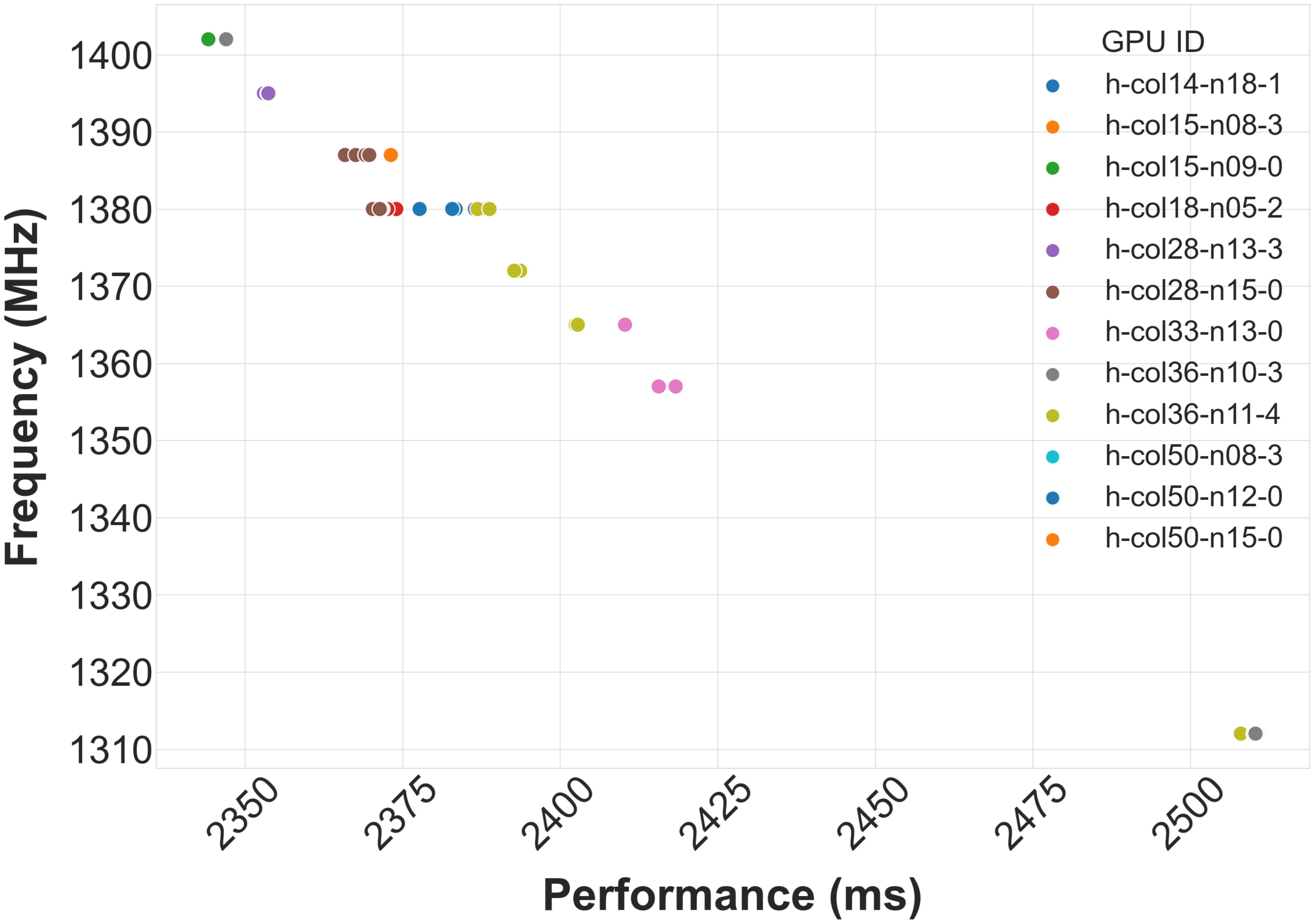}
    \caption{Performance v. Frequency}
    \label{fig:summit-rowh-perf-freq-scatterplot}
  \end{subfigure}
  \begin{subfigure}{.45\textwidth}
    \centering
    \includegraphics[width=\textwidth]{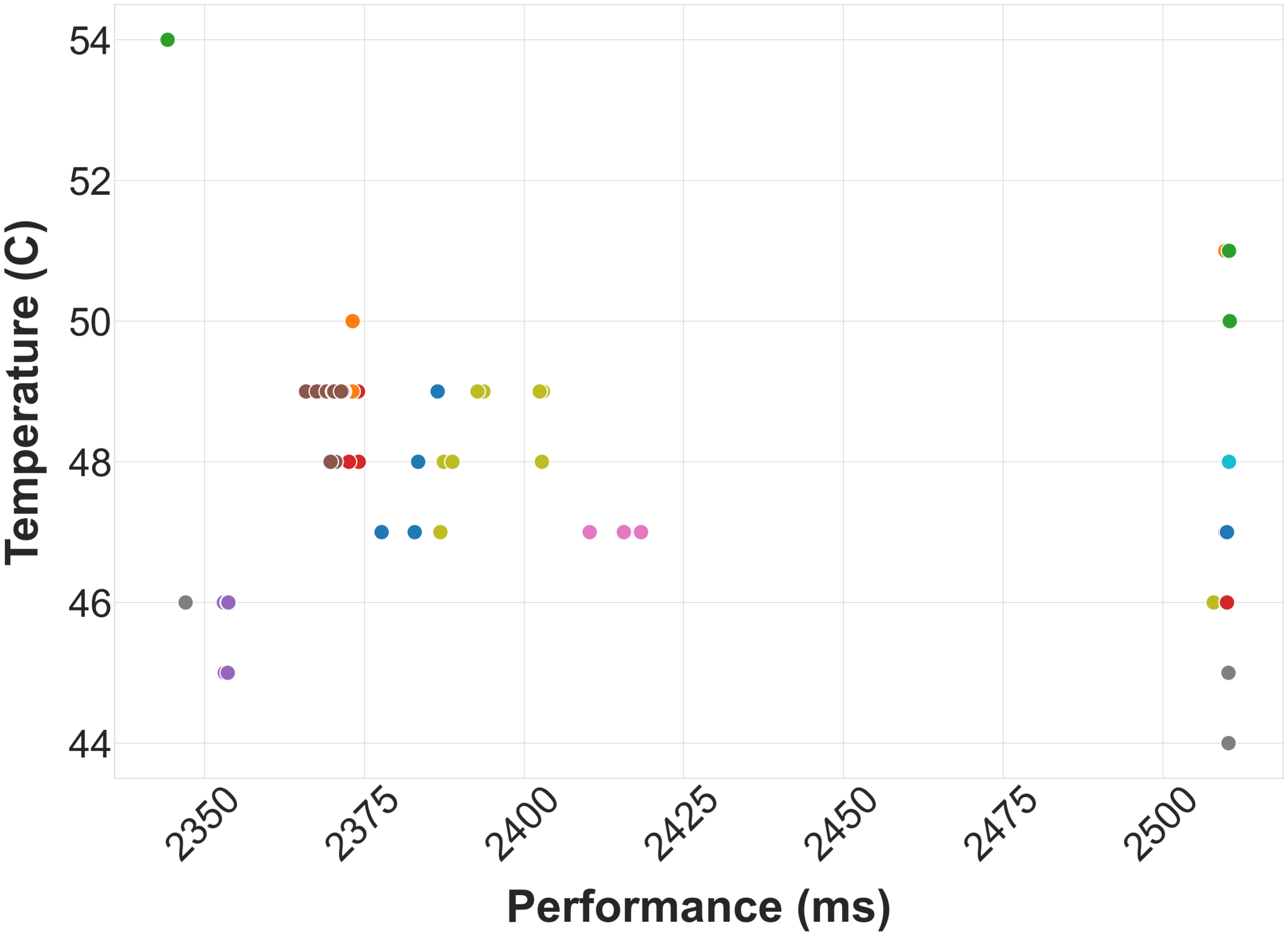}
    \caption{Performance v. Temperature}
    \label{fig:summit-rowh-perf-temp-scatterplot}
  \end{subfigure}
  \begin{subfigure}{.45\textwidth}
    \centering
    \includegraphics[width=\textwidth]{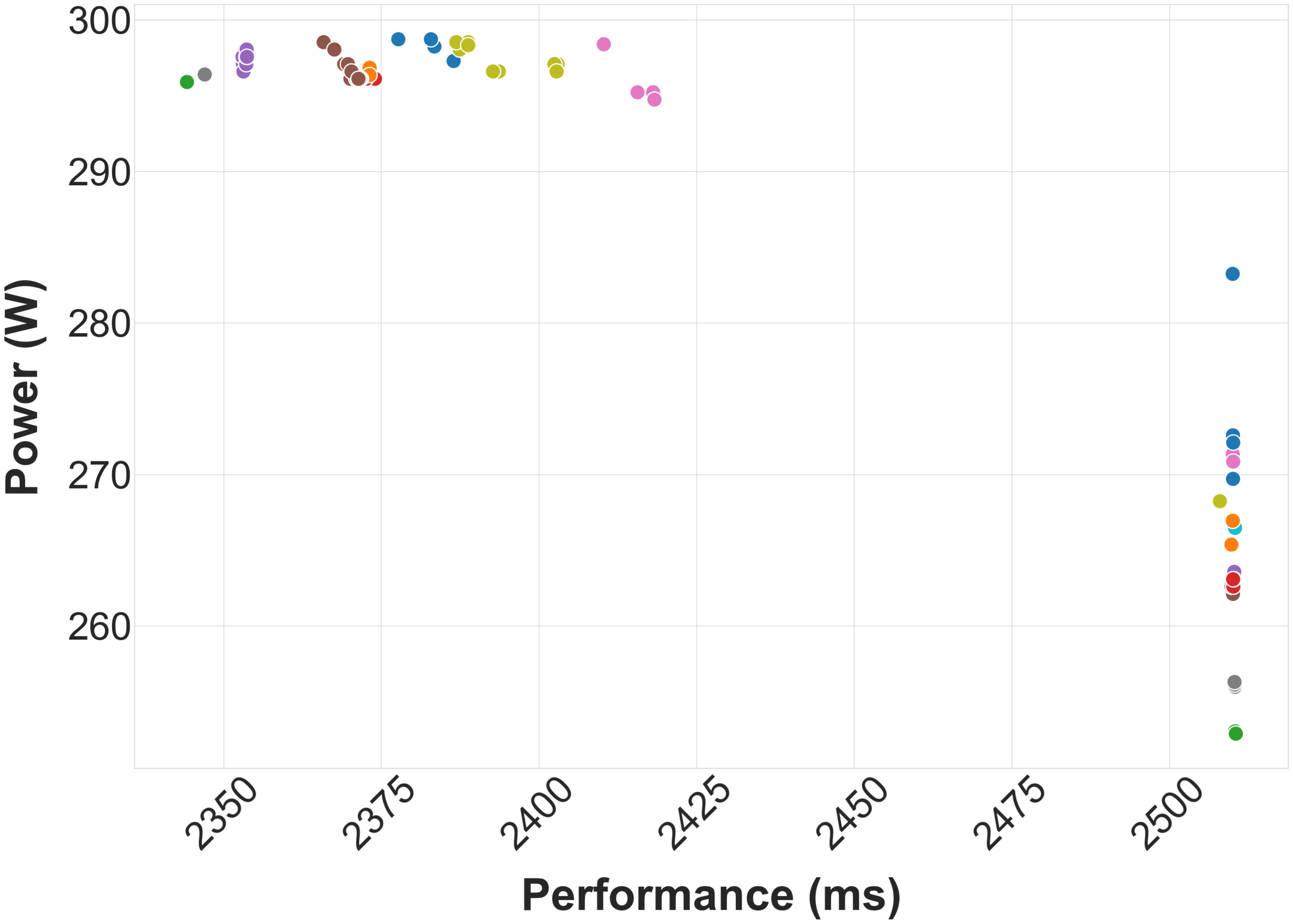}
    \caption{Performance v. Power}
    \label{fig:summit-rowh-perf-pwr-scatterplot}
  \end{subfigure}
  \begin{subfigure}{.45\textwidth}
    \centering
    \includegraphics[width=\textwidth]{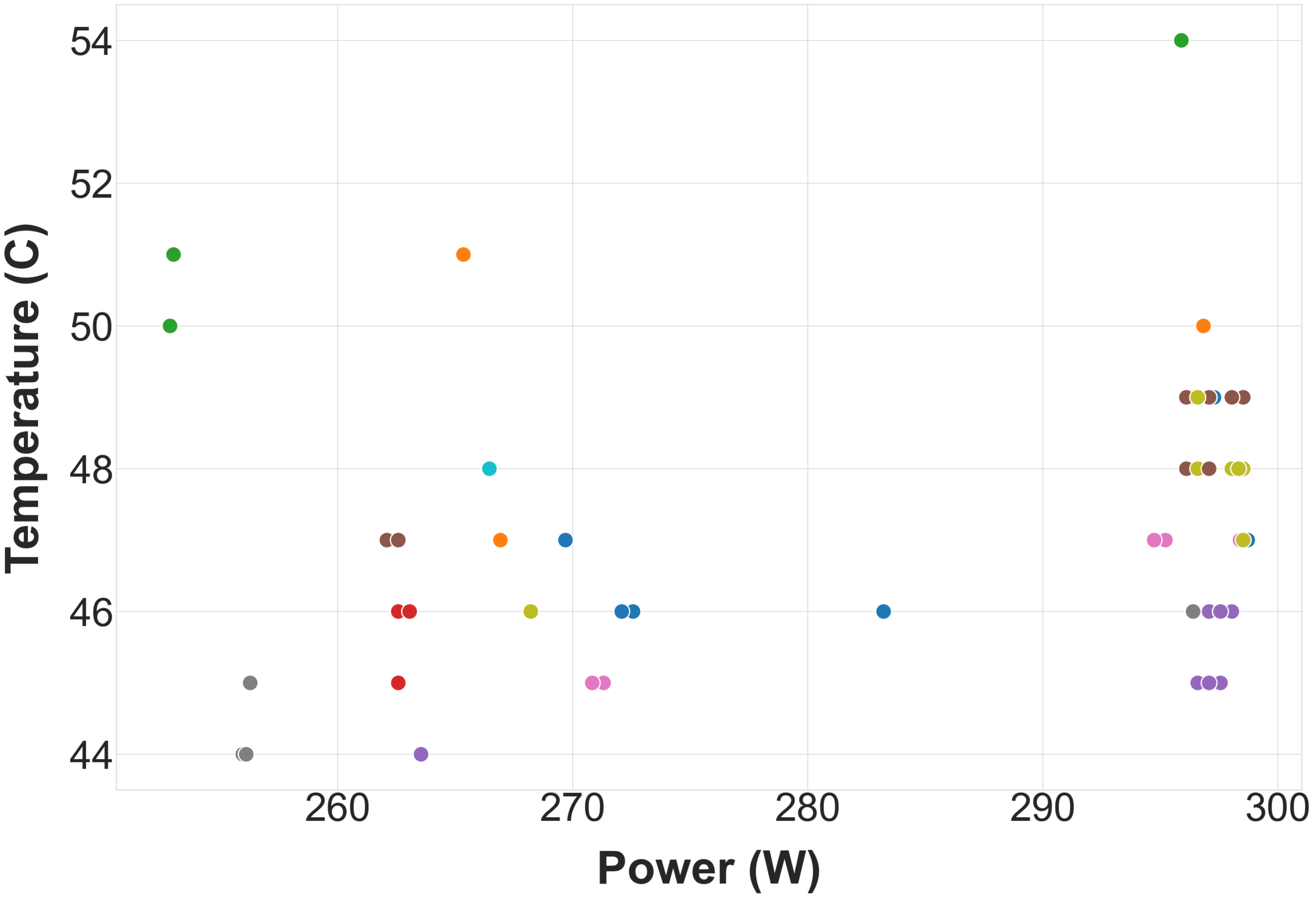}
    \caption{Power v. Temperature}
    \label{fig:summit-rowh-pwr-temp-scatterplot}
  \end{subfigure}
  \vspace{-0.1cm}
  \caption{Scatter plots for GPUs in Row H of Summit showing possible correlations between (a) temperature and performance (kernel duration), (b) frequency and performance, (c) power and performance, and (d) temperature and power. The color indicates the GPU. Only GPUs with at least one reported power level $<$ 290 W are included.}
  \label{fig:summit-row-h-scatterplots}
\end{figure*}

\begin{figure*}[!t]
  \centering
  \begin{subfigure}{\textwidth}
    \centering
    \includegraphics[width=\textwidth]{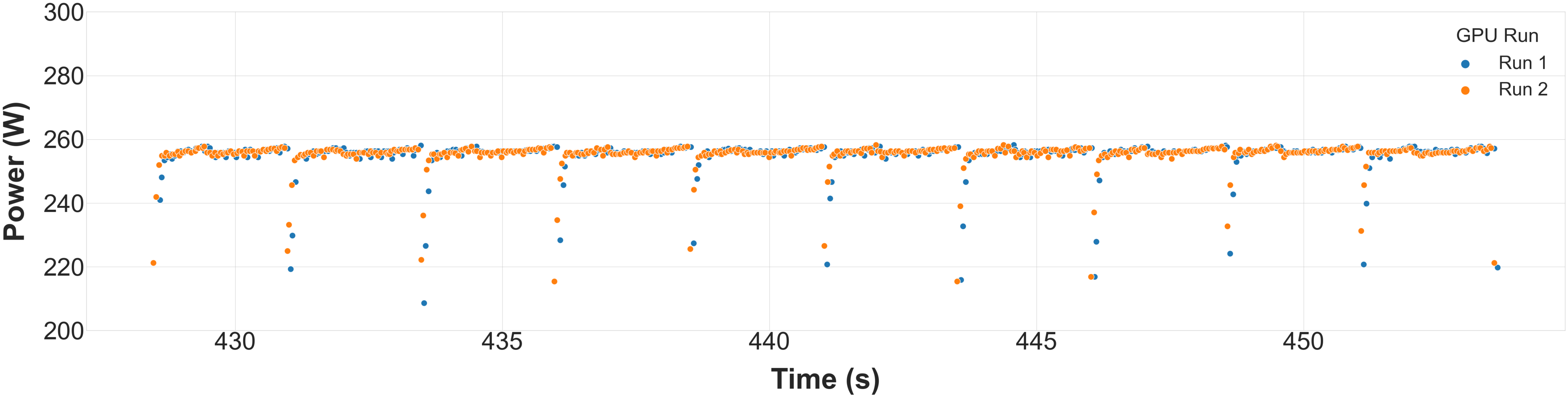}
    \caption{Power}
    \label{fig:summit-rowh-pwr-timeline}
  \end{subfigure}
  \begin{subfigure}{\textwidth}
    \centering
    \includegraphics[width=\textwidth]{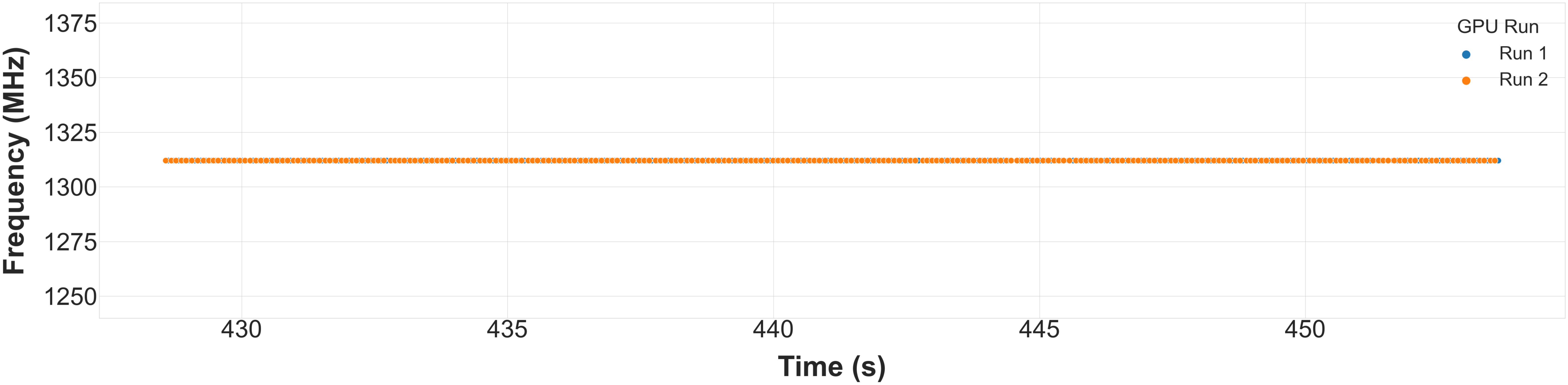}
    \caption{Frequency}
    \label{fig:summit-rowh-freq-timeline}
  \end{subfigure}
  \vspace{-0.3cm}
  \caption{Time-series plots showing continuous measurements for (a) power and (b) frequency for two distinct runs of one GPU in Row H of Summit with at least one reported power level $<$ 290W. The GPU plotted is rowh-col36-n10-3.}
  \label{fig:summit-rowh-timelines}
\end{figure*}

Our results in Section~\ref{subsec:res-summit} show that outliers exist in Summit.
To understand these outliers better, we more deeply examined the results within Summit's row H and found additional interesting trends.
Behavior varies significantly across the different columns in row H, despite being located near one another.
For example, Figure~\ref{fig:summit-row-h-summary} shows that 7 of the 29 columns in row H have no frequency, performance, power, or temperature outliers.
In the remaining columns that do have at least one outlier, the vast majority of the outliers come from columns 13, 14, 28, 33, 36, and 50.
Columns 33 and 36 in particular have a very large number of outliers; both columns have frequency, performance, power, and temperature outliers (the remaining columns have outliers in 1-3 of these categories).

Figure~\ref{fig:summit-row-h-scatterplots} shows that despite several columns having outliers in multiple metrics, these outliers are not always well correlated.
Note that, since we do not have sole access to Summit and are not guaranteed access to a specific node when enqueuing a job, the number of runs for different nodes in the scatter plots ranges from 1 (column50-node08, column50-node12, and column50-node15) to 16 (column28-node13).
Nevertheless, correlating multiple metrics yields further interesting insights.
Perhaps unsurprisingly, performance and frequency (Figure~\ref{fig:summit-rowh-perf-freq-scatterplot}) are often well correlated -- i.e., as frequency in a given GPU decreases, performance also increases.
However, the correlation between performance and temperature (Figure~\ref{fig:summit-rowh-perf-temp-scatterplot}), performance and power (Figure~\ref{fig:summit-rowh-perf-pwr-scatterplot}), and power and temperature (Figure~\ref{fig:summit-rowh-pwr-temp-scatterplot}) show significantly less correlation.
In particular, the power outliers show fascinating trends when correlated with performance: although the outliers all take approximately 2510 ms to complete, their power consumption across the different nodes is very different -- ranging from 250 - 285W, despite running for the same amount of time.
Interestingly, every GPU in the scatter plots appears to have at least one outlier.
For example, rowh-col28-n13-3, shown in purple in Figure~\ref{fig:summit-rowh-perf-pwr-scatterplot}, appears to be representative of a GPU that consistently does not show outliers: it usually completes in 2353 ms while consuming 295 - 299W.
However, it has an outlier where it consumes 264W and completes in 2510 ms.
Rowh-col14-n18-1, shown in blue in Figure~\ref{fig:summit-rowh-perf-pwr-scatterplot}, is more consistent than rowh-col28-n13-3, except that it consistently has outliers: it completes in 2510 ms while consuming 269 - 273W.
Other GPUs show more divergent behavior, including runs that are outliers and runs that are not outliers.
For example, rowh-col36-n11-4, shown in lime green in Figure~\ref{fig:summit-rowh-perf-pwr-scatterplot}, has runs that consume up to 30W different power while taking between 2385 and 2510 ms to complete (even though the range of temperatures for this GPU is less than 3\degree C).
This is perhaps unsurprising, because rowh-col36-n11-4 has wider variance in frequency (68 MHz, Figure~\ref{fig:summit-rowh-perf-freq-scatterplot}\footnote{In Figure~\ref{fig:summit-rowh-perf-freq-scatterplot}, we jitter the yellow point from (2510, 1312) to (2508, 1312) to make the frequency range of 68 MHz for this GPU (rowh-col36-n11-4) easily noticeable.})
than other GPUs.
In particular, this GPU exhibits large swings in frequency, power, and performance -- the water cooling appears to be doing its job since the temperature variation is small, but this is not preventing the GPU from having large variations in behavior.

These interesting results led us to further examine the behavior across multiple runs on one of the same GPUs.
In particular, we selected rowh-col36-node10-3 (shown in gray in Figure~\ref{fig:summit-row-h-scatterplots}), because this GPU exhibits significant variation and contains power outliers.
Figure~\ref{fig:summit-rowh-timelines} also shows a time-series of continuous measurements for power and frequency from two runs on this GPU.
We can see that this GPU clearly exhibits outlier behavior, since the maximum power consumption only reaches 259W.
Even more interestingly, we see that this GPU is always utilizing the same frequency across both runs -- 1312 MHz, despite the instantaneous power consumption rising and falling as the kernels progress.
In particular, the startup power consumption is lower, showing that the GPU is able to increase its power consumption as it reaches the steady state in a given kernel invocation.
However, the frequency consumption does not rise with it, unlike what we observed in TACC.
Nevertheless, given the large number of outliers, across all metrics, row H exemplifies the outliers across Summit.
Yet, because each node in a column of row H contains multiple GPUs, in Appendix~\ref{subsec:res-summit-rowh-col36} we further sub-divide, analyze, and examine the behavior of row H, column 36, since this column exhibits numerous outliers across all metrics.

\subsection{Row H, Column 36 Results}
\label{subsec:res-summit-rowh-col36}

\begin{figure*}[tb!]
  \centering
  \begin{subfigure}{.45\textwidth}
    \centering
    \includegraphics[width=\textwidth]{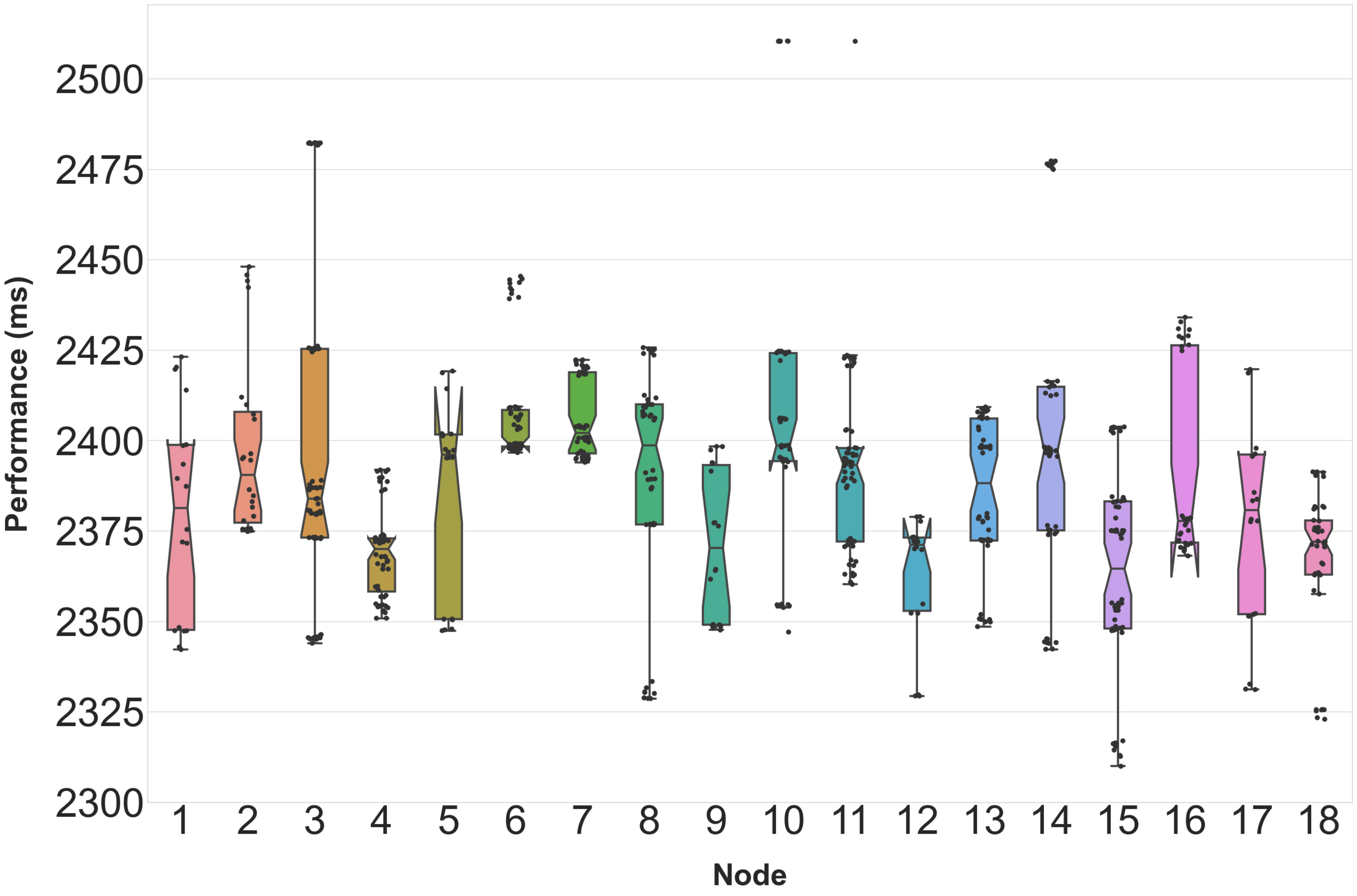}
    \caption{Performance}
    \label{fig:summit-row-h-col-36-perf}
  \end{subfigure}
  \begin{subfigure}{.45\textwidth}
    \centering
    \includegraphics[width=\textwidth]{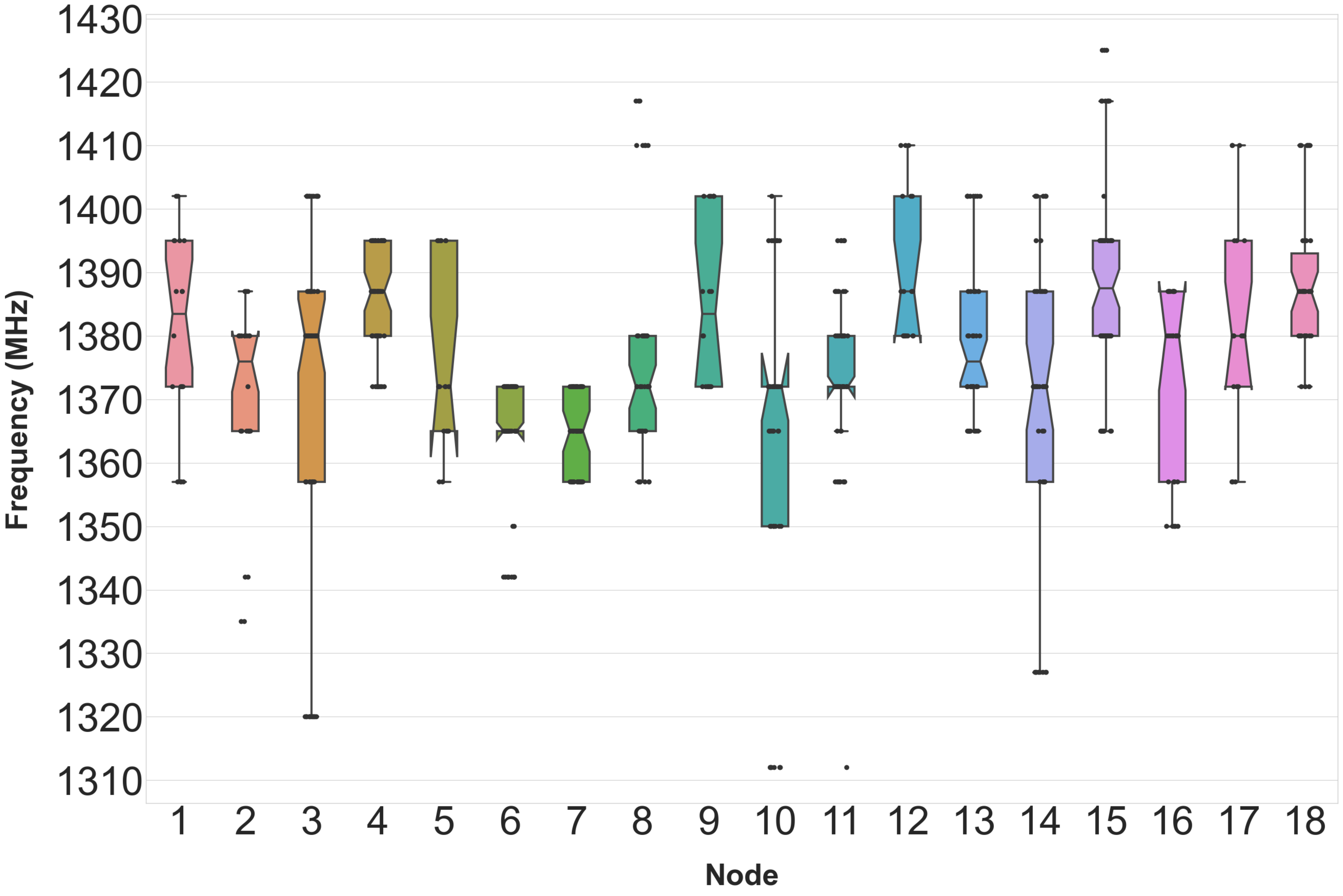}
    \caption{Frequency}
    \label{fig:summit-row-h-col-36-freq}
  \end{subfigure}
  \begin{subfigure}{.45\textwidth}
    \centering
    \includegraphics[width=\textwidth]{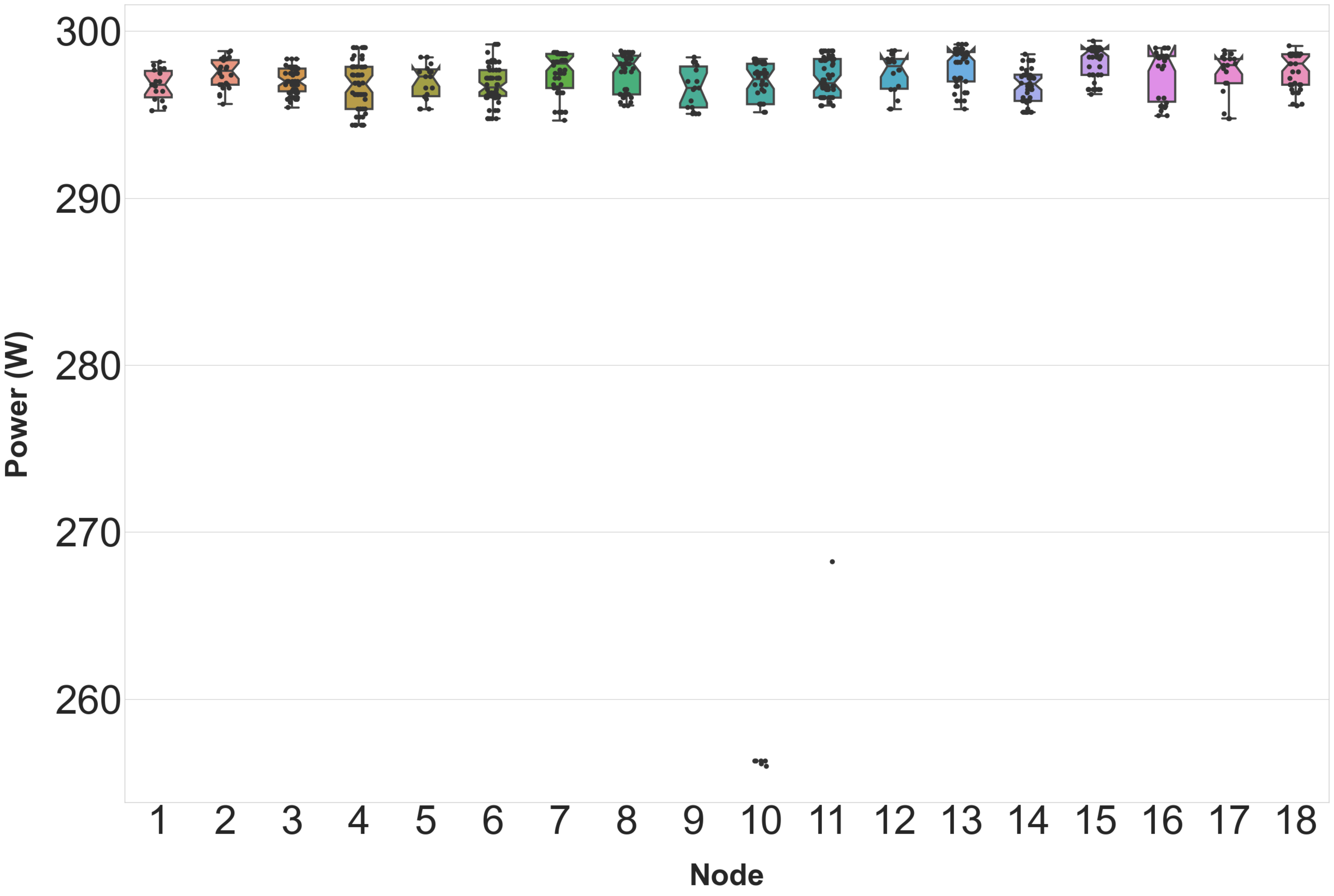}
    \caption{Power}
    \label{fig:summit-row-h-col-36-pwr}
  \end{subfigure}
  \begin{subfigure}{.45\textwidth}
    \centering
    \includegraphics[width=\textwidth]{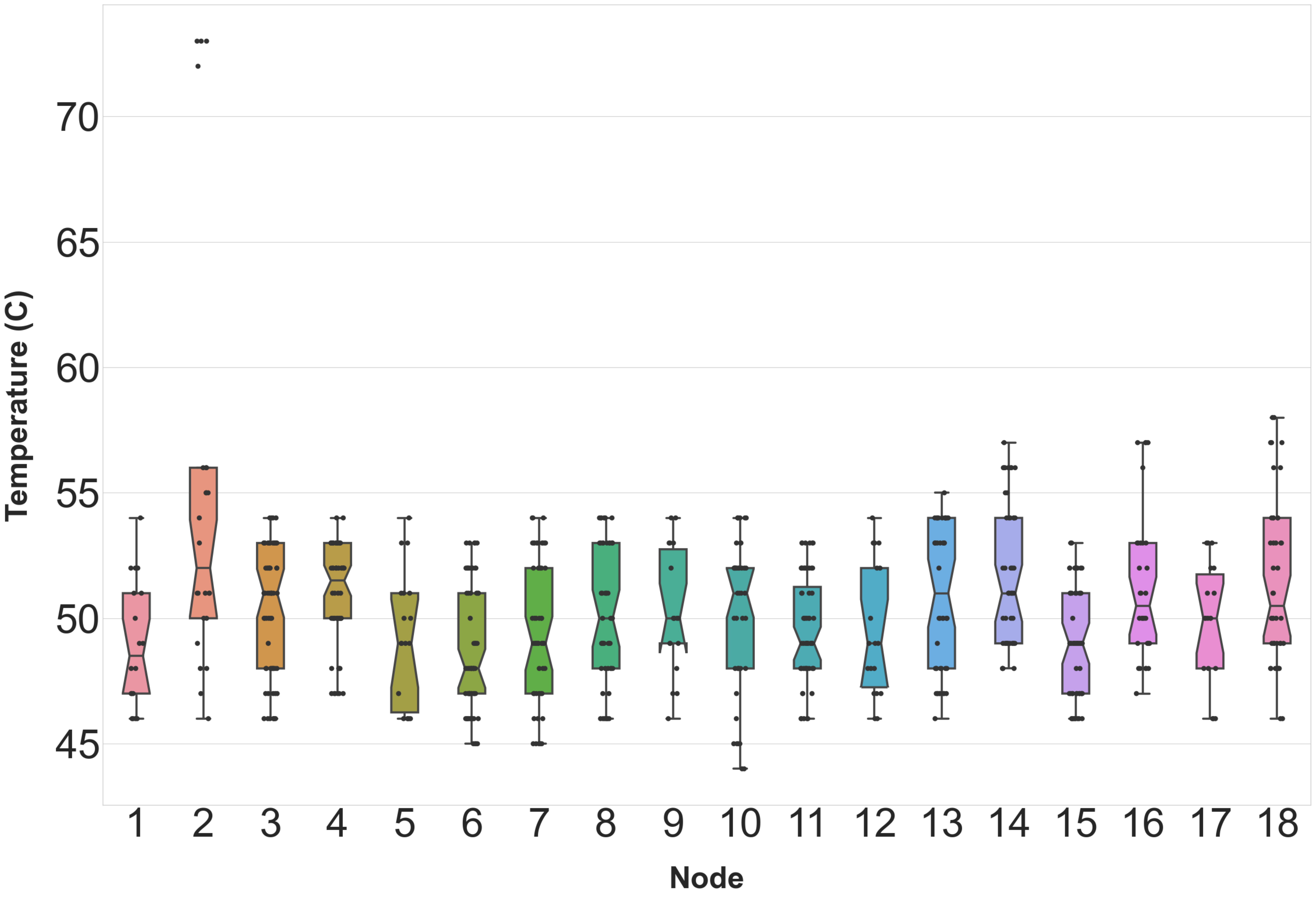}
    \caption{Temperature}
    \label{fig:summit-row-h-col-36-temp}
  \end{subfigure}
  \vspace{-0.3cm}
  \caption{Summary results from Summit Row H Column 36, showing variation in performance (kernel duration), measured frequency, power and temperature as reported by NVIDIA's profiler, when running the GPUs unthrottled at the TDP value of 300W.}
  \label{fig:summit-row-h-col-36-summary}
\end{figure*}

Figure~\ref{fig:summit-row-h-col-36-summary} shows the performance, frequency, power, and temperature breakdown across multiple runs for each node in row H, column 36.
Similar to Appendix~\ref{subsec:res-summit-rowh}, the breakdown within row H, column 36 shows that the outliers come from a specific subset of the nodes: 7 nodes (nodes 2, 6, 8, 10, 11, 13, 14, and 18) have at least one outlier, while the remaining 9 nodes do not have any.
This further reinforces the difficulty in drawing conclusions from the cluster-wide summary, but also highlights the need to intelligently zoom in on specific problematic row-column pairs, given the vast amount of data across the 27,648 GPUs.
In particular, nodes 10 and 11 show the most outliers across frequency, performance, and power.
This is corroborated by Figure~\ref{fig:summit-row-h-scatterplots}, which shows rowh-col36-node10-3 (gray) and rowh-col36-node11-4 (lime green).
These GPUs have a wide disparity in results (including points which are outliers and not outliers), as discussed in Appendix~\ref{subsec:res-summit-rowh}.
In particular, these nodes have large outliers in performance (up to 2510 ms) and power (255W in several cases).
However, neither of these nodes have any temperature outliers, unlike our prior findings on TACC.
Interestingly, all temperature outliers occur on rowh-col36-node2 -- which does not have any frequency, performance, or power outliers, but which runs at up to 73\degree C.
Thus, water cooling does not appear to be completely successful in keeping this node's GPUs within the desired range.
Overall, these results highlight how several nodes repeatedly cause outlier results across several metrics, and do not always fit the established patterns we observed on other clusters.
Like other clusters, frequency is usually a consistent issue for the variations.
However, the results for rowh-col36-node2 show that variations can be caused (in Summit) by temperature alone.
Compared to Vortex, Summit also showed considerably more variation (potentially due to gathering
data across a larger amount of time and larger number of GPUs), despite Summit and Vortex both using water cooling.
Thus, relying on water cooling alone to reduce variation does not appear to be sufficient.

\end{appendices}

\end{document}